\documentclass[11pt]{article}

\usepackage{epsfig}
\usepackage{subfig}
\usepackage{graphicx}
\usepackage{epstopdf}
\usepackage{amsfonts}
\usepackage{amssymb}
\usepackage{amsthm}
\usepackage{amsmath}
\usepackage{multirow}
\usepackage{color}
\usepackage{cite}

\def\beq{\begin{equation}}
\def\eeq{\end{equation}}
\def\bea{\begin{eqnarray}}
\def\eea{\end{eqnarray}}
\newcommand{\beqs}{\begin{subequations}}
\newcommand{\eeqs}{\end{subequations}}

\newcommand{\cref}[1]{Ref.~\cite{#1}}

\newcommand{\bra}[1]{\left<#1\right|}
\newcommand{\ket}[1]{\left|#1\right>}

\newcommand{\hh}{{\ensuremath{I{\kern-2.6pt h}}}}
\newcommand{\bhh}{{\ensuremath{\bar{I{\kern-2.6pt h}}}}}

\def\e{\epsilon}

\def\G{\Gamma}

\def\q2 {q^2}

\def\e{\epsilon}

\def\G{\Gamma}

\def\q2 {q^2}

\def\bra{\langle}
\def\ket{\rangle}

\usepackage[colorlinks=true,allcolors=blue]{hyperref}
\begin{document}

\begin{titlepage}
	
%\vspace*{-15mm}
%\begin{flushright}
%{UT-STPD-21/01}\\
%\end{flushright}
%\vspace*{0.7cm}

\begin{center}
{\Large {\bf A Predictive $SO(10)$ Model}
}
\\[12mm]
George Lazarides,$^{1}$~%\footnote{E-mail: \texttt{lazaride@eng.auth.gr}}
Rinku Maji,$^{2}$~
Rishav Roshan,$^{3}$~
Qaisar Shafi$^{4}$~%\footnote{E-mail: \texttt{shafi@bartol.udel.edu}}
\end{center}
\vspace*{0.50cm}
\centerline{$^{1}$ \it
School of Electrical and
Computer Engineering, Faculty of Engineering,
}

\centerline{\it
Aristotle University
of Thessaloniki, Thessaloniki 54124, Greece}
\vspace*{0.2cm}
	\centerline{$^{2}$ \it
		Theoretical Physics Division, Physical Research Laboratory,}
		\centerline{\it  Navarangpura, Ahmedabad 380009, India}
	\vspace*{0.2cm}
	\centerline{$^{3}$ \it
		Department of Physics, Kyungpook National University, Daegu 41566, Korea}
	%\centerline{\it
	%	 Department of Physics, Kyungpook National University, Daegu 41566, Korea}
	\vspace*{0.2cm}
	\centerline{$^{4}$ \it
		Bartol Research Institute, Department of Physics and 
		Astronomy,}
	\centerline{\it
		 University of Delaware, Newark, DE 19716, USA}
	\vspace*{1.20cm}
\begin{abstract}
We discuss some testable predictions of a non-supersymmetric $SO(10)$ model supplemented by a Peccei-Quinn symmetry. 
We utilize a symmetry breaking pattern of $SO(10)$ that yields unification of the Standard Model gauge couplings, 
with the unification scale also linked to inflation driven by an $SO(10)$ singlet scalar field with a 
Coleman-Weinberg potential. Proton decay mediated by the superheavy gauge bosons may be observable at the proposed 
Hyper-Kamiokande experiment. Due to an unbroken $Z_2$ gauge symmetry from $SO(10)$, the model predicts the presence of a stable intermediate mass fermion which, 
together with the axion, provides the desired relic abundance of dark matter. The model also predicts the presence 
of intermediate scale topologically stable monopoles and strings that survive inflation. The monopoles may be present 
in the Universe at an observable level. We estimate the stochastic gravitational wave background emitted by the strings 
and show that it should be testable in a number of planned and proposed space and land based experiments. Finally, 
we show how the observed baryon asymmetry in the Universe is realized via non-thermal leptogenesis.
\end{abstract}

%\vspace{1.5cm}
%{\Large\textit{\textbf{\textsl{
%\begin{center}
%The Magnetic Monopole Ninety Years Later
%\end{center}
%}}}}
\end{titlepage}
%%%%%%%%%%%%%%%%%%%%%%%%%%%%%%%%%%%%%%%%%%%%%%%%%%%%%%%
\section{Introduction}

A recent paper \cite{Lazarides:2020frf} highlighted some salient features of a non-supersymmetric $SO(10)\times 
U(1)_{\rm PQ}$ model \cite{Holman:1982tb,Mohapatra:1982tc}, where $U(1)_{\rm PQ}$ denotes the Peccei-Quinn (PQ) symmetry included 
to resolve the strong CP problem \cite{Peccei:1977hh,Peccei:1977ur}. The $SO(10)$ symmetry is broken to 
$SU(3)_C\times U(1)_{\rm EM}$ by employing tensor representations such that the $Z_2$ subgroup of $Z_4$, the 
center of $SO(10)$, remains unbroken \cite{Kibble:1982ae}. Independent of the symmetry breaking chain, this 
yields topologically stable cosmic strings  \cite{Kibble:1982ae,Kibble:1982dd} with a string tension that is determined by 
the appropriate symmetry breaking scale. We focus here on a specific symmetry breaking pattern of $SO(10)$ which 
is compatible with the unification of the Standard Model (SM) gauge couplings and also yields topologically stable 
intermediate scale monopoles and strings \cite{Chakrabortty:2020otp,Lazarides:2022spe}. We also take into account 
primordial inflation driven by an $SO(10)\times U(1)_{\rm PQ}$ singlet real scalar field with a 
Coleman-Weinberg potential \cite{Coleman:1973jx,Shafi:1983bd}. In light of the most recent measurements \cite{Akrami:2018odb,BICEP:2021xfz} 
of $n_{\rm s}$ and $r$, the scalar spectral 
index and tensor-to-scalar ratio respectively, a non-minimal coupling of the inflaton to gravity is preferred 
\cite{Maji:2022jzu}. Regarding $U(1)_{\rm PQ}$, we assume that this symmetry is spontaneously broken after 
inflation ends in which case one should make sure that the axion domain wall problem does not exist. This is 
taken care of through the introduction of two fermionic 10-plets whose components acquire masses from the 
breaking of $U(1)_{\rm PQ}$ at scale $f_a$ \cite{Holman:1982tb,Lazarides:1982tw}. An important
consequence of these considerations is the appearance of intermediate scale WIMP-like fermionic dark matter (DM)
whose stability is ensured by the unbroken gauge $Z_2$ symmetry that we previously mentioned.
We are therefore led to a scenario in which the observed dark matter in the universe potentially consists of axions 
as well as electrically neutral intermediate scale fermions from the $SU(2)_L$ doublet components in the 10-plets.
Following Ref.\cite{Lazarides:2020frf} we expand on some of the most important predictions of this $SO(10)\times 
U(1)_{\rm PQ}$ model which includes gauge coupling unification, inflation, proton decay, axion, and heavy WIMP DM, 
and non-thermal leptogenesis implemented within a framework that takes into account the observed fermion 
masses and mixings. 

The paper is organized as follows. In Section~\ref{sec:model} we summarize the salient features of the model including the field content 
and the symmetry breaking pattern. The renormalization group analysis of the SM gauge couplings and proton decay are discussed in 
Section~\ref{sec:RGEs}, and the effects of threshold corrections and dimension-5 operators on the unification of the gauge couplings are 
discussed in Section~\ref{sec:threshold}. In Section~\ref{sec:inflation} we sketch the Coleman-Weinberg inflationary scenario. 
Section~\ref{sec:top} is devoted to the generation of topological defects (monopoles and cosmic strings) and the gravitational wave 
spectrum from the decay of cosmic string loops. In Section~\ref{sec:dm_lepto} we construct the Boltzmann equations for the production of 
the baryon asymmetry of the Universe via non-thermal leptogenesis as well as the non-thermal generation of fermionic DM. In 
Section~\ref{sec:numeric} we analyse the axion contribution to the DM abundance and solve numerically the Boltzmann equations in two 
examples. Our conclusions are summarized in Section~\ref{sec:conclusion}.

%The paper is organized as follows. In Sec.~\ref{sec:model} we summarize the salient features of the model. The renormalization group analysis 
%and proton decay are outlined in Sec.~\ref{sec:RGEs} and the effect of threshold corrections and dimension-5 operators on the unification of 
%the gauge couplings is discussed in Sec.~\ref{sec:threshold}. In Sec.~\ref{sec:inflation} we sketch the Coleman-Weinberg inflationary scenario. 
%Section~\ref{sec:top} is devoted to the generation of topological defects (monopoles and cosmic strings) and the gravitational wave spectrum 
%from the decay of cosmic string loops. In Sec.~\ref{sec:dm_lepto} we construct the Boltzmann equations for the production of the baryon 
%asymmetry of the Universe via non-thermal leptogenesis as well as the non-thermal generation of fermionic DM. In Sec.~\ref{sec:numeric} we 
%analyse the contribution of the axion to DM and solve numerically the Boltzmann equations in two examples. Finally, in 
%Sec.~\ref{sec:conclusion} we summarize our conclusions. 

%%%%%%%%%%%%%%%%%%%%%%%%%%
\section{$SO(10)\times U(1)_{\rm PQ}$ Symmetry Breaking}
\label{sec:model}
The fermion sector consists of three generations of 16-plets and two generations of 10-plets denoted as follows:
\begin{align}
\psi^{(i)}_{16}(1) \ \ (i = 1,2,3)  , \ \  \psi^{(\alpha)}_{10}(-2) \quad (\alpha = 1,2) ,
\end{align}
where the numbers within parentheses are the PQ charges of the respective multiplets.
The complex scalar multiplets are
\begin{align}
\phi_{10}(-2), \ \ \phi_{45}(4), \ \ \phi_{126}(2), \ \ \phi_{210}(0) .
\end{align}

For definiteness, we consider the following breaking scheme 
\begin{align}
\label{chain}
& SO(10)\times U(1)_{\rm PQ}\xrightarrow[M_U]{\left<210(0)\right>} \nonumber \\
& SU(2)_L\otimes SU(2)_R\otimes SU(4)_C\times U(1)_{\rm PQ} \xrightarrow[M_I]{\left<(1,1,15)\in 210(0)\right>} \nonumber \\
& SU(2)_L\otimes SU(2)_R\otimes SU(3)_C \otimes U(1)_{B-L}\times U(1)_{\rm PQ}\xrightarrow[M_{II}]{\left< (1,3,1,-2)\in (1,3,10)\in 
\overline{126}(-2)\right>} \nonumber \\
& SU(3)_C \otimes SU(2)_L\otimes U(1)_Y\otimes\mathbb{Z}_2 \times U(1)'_{\rm PQ}
\xrightarrow[f_a]{\left<(1,3,1)+(1,1,15)\in 45(4)\right>} \nonumber \\
& SU(3)_C \otimes SU(2)_L\otimes U(1)_Y\otimes\mathbb{Z}_2 \xrightarrow[m_W]{\left< (1,2,\pm \frac{1}{2})\in 10(-2)\right>} SU(3)_C \otimes U(1)_Q\otimes\mathbb{Z}_2.
\end{align}
%%%%%%%%%%%%%%%%%%%%%%%%%%%%%%%%%%%%%%%%%%%
{
%\setstretch{1.5}
\begin{table}[htbp]
\begin{center}
\begin{tabular}{| c | c | c | c | c |}
\hline
& $SO(10)\times U(1)_{PQ}$ & $\mathcal{G}_{2_L2_R4_C}\times U(1)_{PQ}$ & $\mathcal{G}_{2_L2_R3_C1_{B-L}}\times U(1)_{PQ}$ & $\mathcal{G}_{3_C2_L1_Y\mathbb{Z}_2} \times U(1)'_{PQ}$ \\
\hline
 \parbox[t]{5mm}{\multirow{15}{*}{\rotatebox[origin=c]{90}{Scalars}}} & $210(0)$ & $\left<(1,1,1)\right>$ & & \\
& &  ${\bf (1, 1, 15)}$ & $\left<(1,1,1,0)\right>$ & \\
& & & $(1,1,3,\frac{4}{3})_{\rm GB}$ & \\
& & & $(1,1,\overline{3},-\frac{4}{3})_{\rm GB}$ & \\
& & & $(1,1,8,0)$ & \\
& & $(1, 3, 15)$ & & \\
& & $(3, 1, 15)$ & & \\
& & $(2, 2, 6)_{\rm GB}$ & & \\
& & $(2, 2, 10)$ & & \\
& & $(2, 2, \overline{10})$ & & \\
& $\overline{126}(-2)$ & ${\bf (1,3,10)}$ & ${\bf (1,3,1,-2)}$ & $\left<(1,1,0)\right>$ \\
& & & & $(1,1,-1)_{\rm GB}$ \\
& & & & $(1,1,-2)$ \\
& & & $(1,3,3,-\frac{2}{3})$ & \\
& & & $(1,3,6,\frac{2}{3})$ & \\
& & ${\bf (2,2,15)}$ & ${\bf (2,2,1,0)}$ & $(1,2,\pm \frac{1}{2})$ \\
& & & $(2,2,3,\frac{4}{3})$ & \\
& & & $(2,2,\overline{3},-\frac{4}{3})$ & \\
& & & $(2,2,8,0)$ & \\
& & $(1, 1, 6)$ & & \\
& & $(3,1,\overline{10})$ & & \\
& $45(4)$ & ${\bf (1,1,15)}$ & $(1,1,1,0)$ & \\
& & & $(1,1,3,\frac{4}{3})$ & \\
& & & $(1,1,\overline{3},-\frac{4}{3})$ & \\
& & & $(1,1,8,0)$ & \\
& & ${\bf (1,3,1)}$ & ${\bf (1,3,1,0)}$ & $(1,1,0)$ \\
& & & & $(1,1,\pm{1})$ \\
& & $(3, 1, 1)$ & & \\
& & $(2, 2, 6)$ & & \\
& $10(-2)$ & ${\bf (2,2,1)}$ & ${\bf (2,2,1,0)}$ & $(1,2,\pm \frac{1}{2})$ \\
& & $(1, 1, 6)$ & & \\
\hline
% \parbox[t]{5mm}{\multirow{3}{*}{\rotatebox[origin=c]{90}{DM}}} & $10(-2)$ & $(1,1,6)$ & $(1,1,3,-\frac{2}{3})$ & $(1,3,-\frac{1}{3})$ \\
% & & & $(1,1,\overline{3},\frac{2}{3})$ & $(1,\overline{3},\frac{1}{3})$\\
% & & $(2,2,1)$ & $(2,2,1,0)$ & $(1,2,\pm \frac{1}{2})$ \\
%\hline
\end{tabular}
\caption{Representations of scalar multiplets at different stages of gauge symmetry breaking. We denote the gauge symmetry by the 
subscripts of the caligraphy $\mathcal{G}$. The numbers represent the dimension of the multiplets under the non-Abelian gauge 
groups along with the charges under the Abelian gauge groups. The multiplets in bold fonts are those that remain massless and 
contribute to the Renormalization Group Equations (RGEs) of gauge couplings whereas the ones with the subscript GB are Goldstone 
bosons eaten by the gauge fields. The rest of the multiplets are integrated out at the breaking scale of the parent gauge symmetry. 
We keep one linear combination of the four SM doublets light after the breaking of the left-right symmetry at $M_{II}$.}
\label{tab:RGE_multiplets}
\end{center}
\end{table}
}
%%%%%%%%%%%%%%%%%%%%%%%%%%%%%%%%%%%%%%%%%%%
Here $M_U$, $M_I$, and $M_{II}$ respectively denote the grand unification and the two intermediate gauge symmetry breaking
scales and $f_a$ is the breaking scale of $U(1)'_{\rm PQ}$. The representations of the multiplets that remain massless at
different stages of gauge symmetry breaking are shown in Table~\ref{tab:RGE_multiplets}. The vacuum expectation value (VEV) 
of $210(0)$ along the $(1,1,1)$ direction breaks $SO(10)$ to the Pati-Salam (PS) gauge group $\mathcal{G}_{2_L2_R4_C}$ \cite{Pati:1974yy} at the unification scale $M_U$. At this stage, $(1,1,15)\in 210$, $(1,3,10)$ and $(2,2,15)$ from $\overline{126}(-2)$, $(1,3,1)$ and $(1,1,15)$ from $45(4)$, 
and the bi-doublet from $10(-2)$ remain massless. The breaking at $M_I$ of $\mathcal{G}_{2_L2_R4_C}$ to $\mathcal{G}_{2_L2_R3_C1_{B-L}}$ 
is achieved via the VEV of the appropriate component of $(1,1,15)\in 210$. At $M_{II}$, a VEV along the SM-singlet direction in 
$(1,3,1,-2)\in \overline{126}(2)$ breaks $\mathcal{G}_{2_L2_R3_C1_{B-L}}$ to $SU(3)_C\times SU(2)_L\times U(1)_Y$, leaving in addition
an unbroken $Z_2$ which is the subgroup of the center $Z_4$ of $Spin(10)$ \cite{Kibble:1982ae}. Note that the $U(1)_{\rm PQ}$ symmetry, so 
far unbroken, is rotated to another global anomalous $U(1)_{\rm PQ}'$ symmetry generated by $Q'_{\rm PQ} = 5Q_{\rm PQ} - 3(B-L) + 4 T_R^3$, 
where $T_R^3$ is the diagonal generator of $SU(2)_R$. The VEV of $45(4)$ finally breaks the $U(1)_{\rm PQ}'$ symmetry at the scale $f_a$.

%%%%%%%%%%%%%%%%%%%%%%%%%%%%%%%%%%%%%%%%%%%%%%%%%%%%%%%%
\section{Renormalization Group Evolution and Proton Decay}
\label{sec:RGEs}
The renormalization group evolution of the gauge couplings $g_i$ ($i=1,2,...,n$) in a generic product gauge group of the form $\mathcal{G} 
\equiv\mathcal{G}_1\otimes \mathcal{G}_2\otimes...\otimes \mathcal{G}_n$ containing non-Abelian groups and at most a single Abelian group 
is governed by the equations \cite{PhysRevLett.30.1343,Caswell:1974gg,Jones:1974mm,Langacker:1980js,Jones:1981we,Slansky:1981yr,Machacek:1983tz,Machacek:1983fi,Machacek:1984zw}:
\begin{equation}
\mu \frac{dg_i}{d\mu} = \frac{1}{16\pi^2} b_i g_i^3 + \frac{1}{(16\pi^2)^2}\sum_{j=1}^n b_{ij} g_i^3g_j^2,
\end{equation}
where $\mu$ is the renormalization scale parameter and
\begin{eqnarray}\label{beta_coef}
b_i &= &\frac{4}{3}\kappa T(F_i)D_{F_i}
+\frac{1}{3}\eta T(S_i)D_{S_i} - \frac{11}{3} C_2(\mathcal{G}_i),  \nonumber \\
b_{ij} & = &\left(\frac{20}{3} C_2(\mathcal{G}_i)\delta_{ij}+4C_2(F_j)\right)\kappa T(F_i)D_{F_i} \nonumber  \\
& &+ \left(\frac{2}{3} C_2(\mathcal{G}_i)\delta_{ij}+4C_2(S_j)\right)\eta T(S_i)D_{S_i} - \frac{34}{3} (C_2(\mathcal{G}_i))^2\delta_{ij} 
\end{eqnarray}
are the one- and two-loop $\beta$-coefficients respectively. The representation of a field multiplet is denoted as $R=(R_1,R_2, ... , R_n)$ 
where $R\equiv F$ for fermions and $R\equiv S$ for scalars. Here, $\kappa=1\;(1/2)$ for Dirac (Weyl) fermions, $\eta=1\;(1/2)$ for complex 
(real) scalars, $T(R_i)$ is the normalization of the representation $R_i$, $C_2(\mathcal{G}_i)$ is the quadratic Casimir operator for the 
group $\mathcal{G}_i$, and $C_2(R_i)$ is the quadratic Casimir operator for the representation $R_i$. Also, 
$D_{R_i}=\prod_{j\neq i}D(R_j)$ with $D(R_i)$ being the dimension of the $i$th representation in the multiplet. For an Abelian group 
$\mathcal{G}_i=U(1)_i$ and a representation $R_i$ with charge $q_i$, we set $T(R_i) = C_2(R_i)=q_i^2$ and $C_2(\mathcal{G}_i)=0$.
%%%%%%%%%%%%%%%%%%%%%%%%%%%%%%%%%%%%%%%%%%%%%%%

  The \textit{extended survival hypothesis} (ESH) \cite{delAguila:1980qag} states that at the level of unbroken gauge symmetry, the only scalars that remain light  
are the ones required to provide the VEVs for breaking this and the subsequent gauge symmetries. This hypothesis provides a prescription for choosing a 
minimal scalar sector at any stage of gauge symmetries.  We use the ESH to choose the scalar sector of the model as given in Table~\ref{tab:RGE_multiplets}. The multiplets in bold fonts are those that remain massless at each stage of 
symmetry breaking and contribute to the RGEs of the gauge couplings. The rest of the multiplets are heavy and decoupled at the parent 
gauge symmetry breaking scale. We keep one linear combination of the four SM doublets to be light after the breaking of the 
$\mathcal{G}_{2_L2_R3_{C}1_{B-L}}$ symmetry at $M_{II}$. Table~\ref{tab:RGE_beta_coef} shows the one- and two-loop beta coefficients 
for the renormalization group evolution of the gauge couplings at different stages of gauge symmetry starting from the scale $M_U$ to 
the mass $m_{\rm DM}$ of the fermionic DM particles.
%%%%%%%%%%%%%%%%%%%%%%%%%%%%%%%%%%%%%%%%
\begin{table}[htb!]
\begin{center}
\begin{tabular}{| c | c | c |}
\hline
 $\mathcal{G}_{2_L2_R4_C}\times U(1)_{PQ}$ & $\mathcal{G}_{2_L2_R3_C1_{B-L}}\times U(1)_{PQ}$ & $\mathcal{G}_{3_C2_L1_Y\mathbb{Z}_2} \times U(1)'_{PQ}$ \\
 \hline
$\begin{pmatrix}
\frac{10}{3} \\ 
\frac{32}{3} \\ 
1
\end{pmatrix}$ , 
$\begin{pmatrix}
\frac{268}{3} & 51 & \frac{525}{2} \\ 
51 & \frac{884}{3} & \frac{1245}{2} \\ 
\frac{105}{2} & \frac{249}{2} & \frac{1109}{2}
\end{pmatrix}$ & $\begin{pmatrix}
-\frac{4}{3} \\ 
0 \\ 
-\frac{17}{3} \\ 
\frac{41}{6}
\end{pmatrix}$ , 
$\begin{pmatrix}
\frac{86}{3} & 9 & 12 & \frac{3}{2} \\ 
9 & 66 & 12 & \frac{27}{2} \\ 
\frac{9}{2} & \frac{9}{2} & -\frac{2}{3} & \frac{7}{6} \\ 
\frac{9}{2} & \frac{81}{2} & \frac{28}{3} & \frac{187}{6}
\end{pmatrix}$ & $\begin{pmatrix}
-\frac{17}{3} \\ 
-\frac{11}{6} \\ 
\frac{163}{30}
\end{pmatrix}$ , 
$\begin{pmatrix}
-\frac{2}{3} & \frac{9}{2} & \frac{41}{30} \\ 
12 & \frac{133}{6} & \frac{3}{2} \\ 
\frac{164}{15} & \frac{9}{2} & \frac{667}{150}
\end{pmatrix}$\\
\hline
\end{tabular}
\caption{One- and two-loop beta coefficients for the renormalization group evolution of the gauge couplings at different stages of gauge symmetry. The light scalar multiplets that contribute to the RGEs are listed in bold fonts in Table~\ref{tab:RGE_multiplets}.}\label{tab:RGE_beta_coef}
\end{center}
\end{table}

 The dimension-6 operators that mediate the decay $p\to \pi^0 e^+$ are given in the physical basis as \cite{Weinberg:1979sa, Wilczek:1979hc, Weinberg:1980bf, Abbott:1980zj,FileviezPerez:2004hn,Nath:2006ut}
\begin{align}
    \label{eq:operator_physical_basis}
    \mathcal{O}_L\left( e^c, d\right) = \mathcal{W}_C\;  \epsilon^{ijk} \overline{u^c_i}\gamma^\mu u_j \overline{e^c} \gamma_\mu d_{k} , \;\;\;
    \mathcal{O}_R\left( e, d^c\right) = \mathcal{W}_C \; \epsilon^{ijk}
    \overline{u^c_i}\gamma^\mu u_j \overline{d^c_{k}} \gamma_\mu e ,
\end{align}
where the Wilson coefficient $\mathcal{W}_C = \frac{g_U^2}{2 M_U^2} \left[ 1 +
|V_{ud}|^2\right]$, with $|V_{ud}| = 0.9742$ being the CKM matrix element \cite{Tanabashi:2018oca}.

The partial lifetime for the $p\to \pi^0 e^+$ channel is expressed as:
\begin{align}
   \tau_p& =\left(\frac{m_p}{32\pi}\left(1-\frac{m_{\pi^0}^2}{m_p^2}\right)^2 R_L^2 \frac{g_U^4}{4 M_U^4}(1+|V_{ud}|^2)^2
 \left( R_{SR}^2 |\langle \pi^0 \rvert (ud)_R u_L\lvert p \rangle |^2 +
     R\to L
 \right)\right)^{-1},
\end{align}
where $m_p$ and $m_{\pi^0}$ denote the proton and pion masses
respectively. $R_L$ is the long-range renormalization factor for the
proton decay operator from the electroweak scale $(m_Z)$ to the QCD
scale ($\sim 1$ GeV) \cite{Nihei:1994tx}, and $R_{SR(SL)}$ is the short-range enhancement factor arising from the renormalization group evolution of
the proton decay operator $\mathcal{O}_{R(L)}$ from $M_U$ to $m_Z$ \cite{Buras:1977yy}. The short-range enhancement factors depend on the breaking 
chain and can be written in the presence of multiple intermediate scales as \cite{Buras:1977yy,Goldman:1980ah,Caswell:1982fx,DANIEL1983219,Ibanez:1984ni,MUNOZ198655}: 
\begin{align} \label{eq:short_range_re_factor}
R_S = \prod_{j}^{M_U\geq M_{j}> m_Z} \prod_i \left[ \frac{\alpha_i \left(M_{j+1}\right)}{\alpha_i \left(M_{j}\right)} \right]^{\frac{\gamma_i}{b_i}} ,
\end{align}
where $\gamma_i$'s are the anomalous dimensions given in Table~\ref{table_anomalous} and $b_i$'s are the one-loop $\beta$-coefficients at different 
stages of the renormalization group evolution from the scale $M_{j}$ to the next smaller scale $M_{j+1}$ (see Table \ref{tab:RGE_beta_coef}).
%%%%%%%%%%%%%%%%%%%%%%%%%%%%%%%%%%%%%%%%%%%%%%%%%%%%%%%%%%%
\begin{table}[htb!]
	\begin{center}
		\begin{tabular}{|c|c|c|}
			\hline
			\multirow{2}{*}{Gauge group} & \multicolumn{2}{c|}{Anomalous dimensions}\\
			\cline{2-3}
			& $\mathcal{O}^{d=6}_L\left(e^c , d\right)$ & $\mathcal{O}^{d=6}_R\left(e , d^c \right)$  \\ 
			\hline
			$\mathcal{G}_{2_L 2_R 4_C}$& $\lbrace \frac{9}{4} , \frac{9}{4} , \frac{15}{4}
			\rbrace$ & $\lbrace \frac{9}{4} , \frac{9}{4} , \frac{15}{4}
		  \rbrace$  \\
			\hline
			$\mathcal{G}_{2_L 2_R 3_C 1_{B-L}}$& $\lbrace \frac{9}{4} , \frac{9}{4} , 2 , \frac{1}{4} \rbrace$ & $\lbrace \frac{9}{4} , \frac{9}{4} , 2 , \frac{1}{4} \rbrace$ \\
			\hline
			$\mathcal{G}_{3_C 2_L 1_Y}$& $\lbrace \frac{9}{4} , \frac{23}{20} , 2
			\rbrace$ &  $\lbrace \frac{9}{4} , \frac{11}{20} , 2 
			\rbrace$ \\
			\hline
		\end{tabular}
		\caption{Relevant anomalous dimensions corresponding to the successive unbroken gauge groups of the considered breaking chain.}
		\label{table_anomalous}
	\end{center}
	
\end{table}

%%%%%%%%%%%%%%%%%%%%%%%%%%%%%%%%%%%%%%%%%%%%%%%%%%%%%%%%%%%%%%%%%%
\section{Unification Solutions and the Effect of Threshold Correction and Dimension-5 Operators}
\label{sec:threshold}
%%%%%%%%%%%%%%%%%%%%%%%%%%%%%%%%%%%%%%%%%%%%%%%%%%%%%%%%%%%%%%
The matching condition when a simple non-Abelian parent group $\mathcal{G}_{\rm P}$ is broken at scale $M_b$ to a subgroup containing a 
non-Abelian factor $\mathcal{G}_{\rm D}$ is given as \cite{WEINBERG198051, Hall:1980kf,Bertolini:2009qj,Bertolini:2013vta, Chakrabortty:2019fov}:
\begin{align}
\frac{1}{\alpha_{\rm D}(M_b)}-\frac{C_2(\mathcal{G}_{\rm D})}{12\pi} &= \frac{1}{\alpha_{\rm P}(M_b)}-\frac{C_2(\mathcal{G}_{\rm P})}{12\pi} 
- \frac{\lambda_{\rm D}(M_b)}{12\pi} \ .
\end{align}
Here $C_2$ represents the quadratic Casimir operator and $\lambda_{\rm D}(M_b)$ is the one-loop threshold correction given by
\begin{align}
\lambda_{\rm D}(M_b)=-21 {\mathrm{Tr}} (t_{DV}^2 \log\frac{m_V}{M_b})+2\eta {\mathrm{Tr}} (t_{DS}^2 \log\frac{m_S}{M_b}) + 
8\kappa \; {\mathrm{Tr}} (t_{DF}^2 \log\frac{m_F}{M_b}),
\end{align}
where $m$ denotes the mass of the heavy fields with the subscripts $V$, $S$, and $F$ for vector, scalar, and fermion states respectively, and 
$t_{D}$ are the generators in the corresponding representation under the daughter symmetry $\mathcal{G}_{\rm D}$. For an Abelian factor in the daughter symmetry, we should 
take $C_2 = 0$ and the trace of the generator squared $t_D^2$ should be
replaced by the corresponding Abelian charge squared. The expressions for the threshold corrections for the various symmetry-breaking scales are given in the Appendix.
%%%%%%%%%%%%%%%%%%%%%%%%%%%%%%%%%%%%%%%%%%%%%%%%%%%%%%%%%%%%%%%%%%
\begin{figure}[htbp!]
\centering
\subfloat[]{\includegraphics[width=0.48\linewidth]{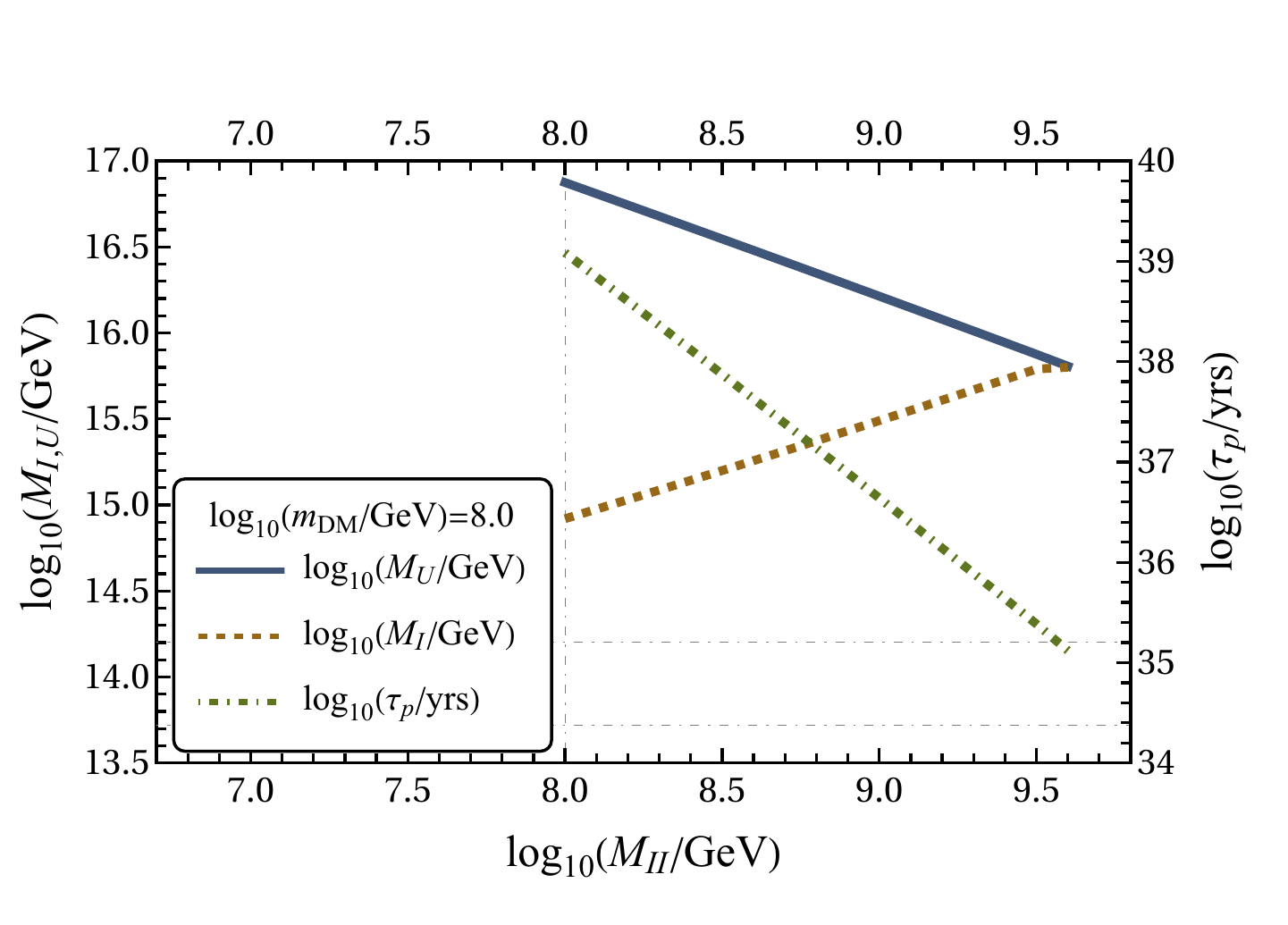}} \hspace{4mm}
\subfloat[]{\includegraphics[width=0.48\linewidth]{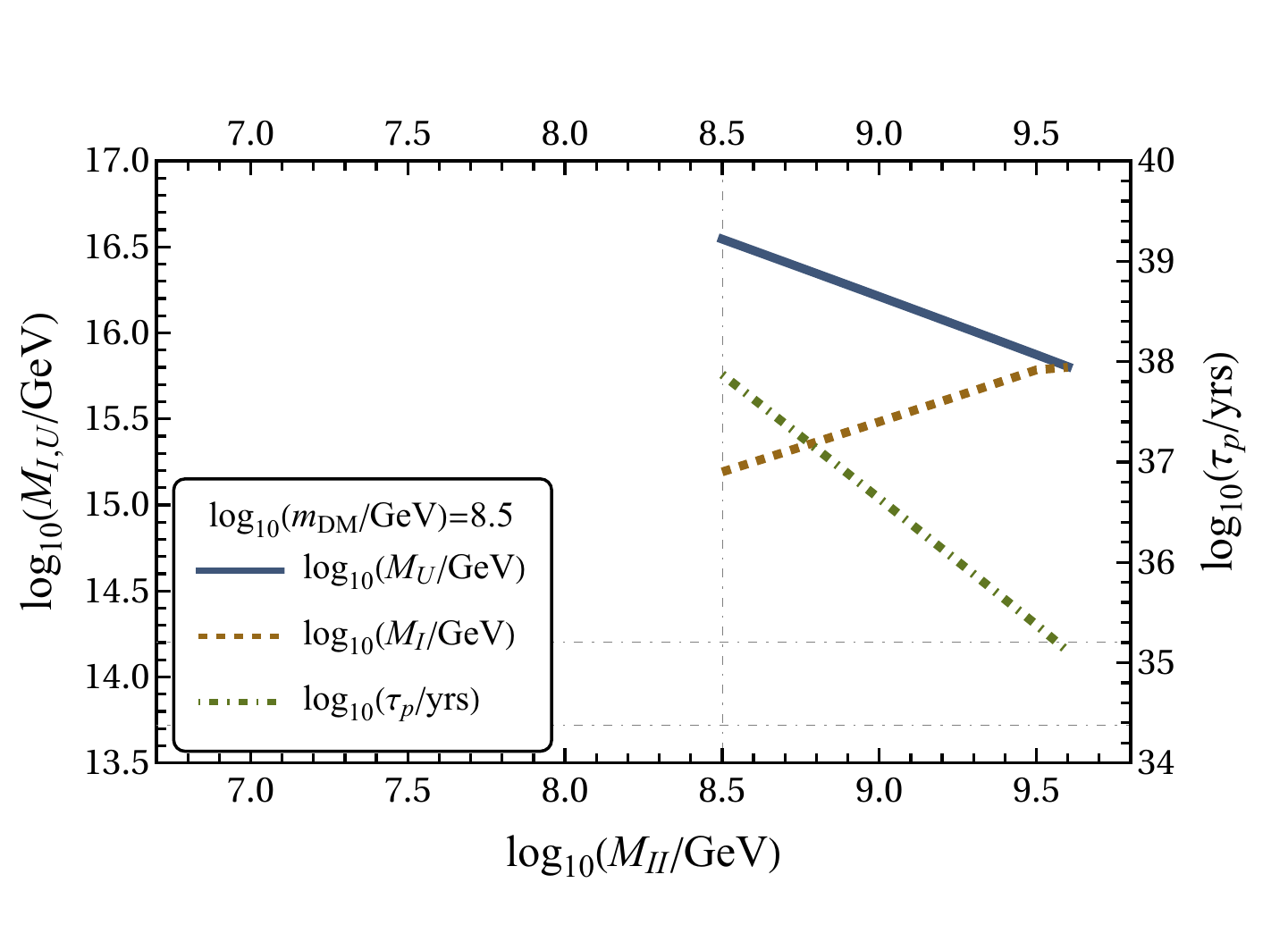}}\\
\subfloat[]{\includegraphics[width=0.48\linewidth]{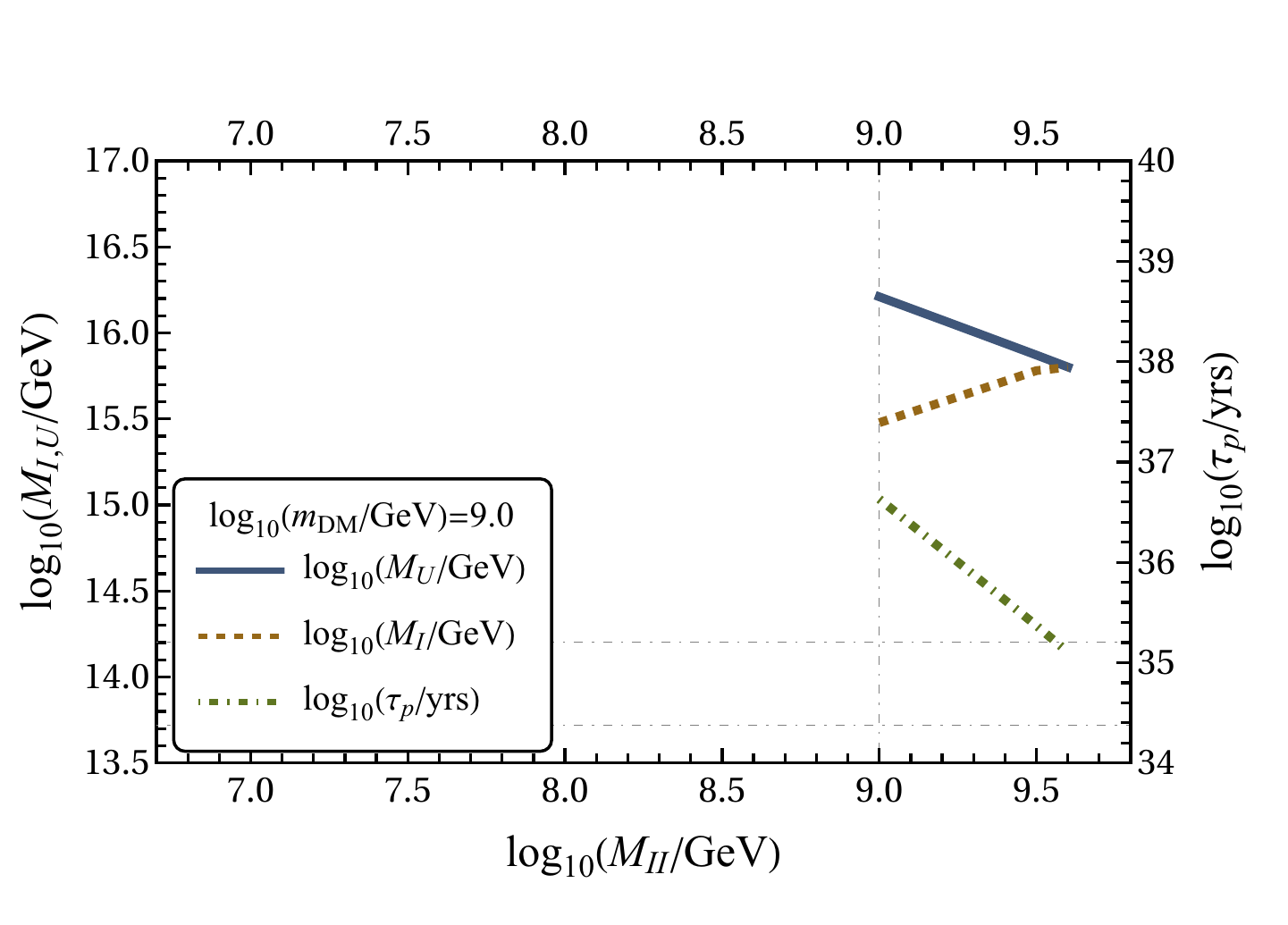}} 
%\hspace{4mm}
%\subfloat[]{\includegraphics[width=0.48\linewidth]{plt7.pdf}}\\
%\subfloat[]{\includegraphics[width=0.48\linewidth]{plt9.pdf}} \hspace{4mm}
%\subfloat[]{\includegraphics[width=0.48\linewidth]{plt11.pdf}}
\caption{Plots of the unification scale $M_U$ and the first intermediate breaking scale $M_I$ along the $Y_1$-axis and the 
partial proton lifetime $\tau_p$ for the channel $p\to \pi^0 e^+$ along the $Y_2$-axis as functions of the second intermediate 
breaking scale $M_{II}$ for different choices of the fermionic DM mass $m_{\rm DM}$ in the range $[10^8,10^{9}]$ GeV. The 
horizontal dash-dotted lines correspond to the present Super-Kamiokande \cite{Super-Kamiokande:2020wjk} and the projected 
Hyper-Kamiokande \cite{Dealtry:2019ldr} limits on $\tau_p$. We see that the proton lifetime is only marginally within the 
reach of the future Hyper-Kamiokande experiment for a narrow range of $M_{II}$.}\label{fig:unific_soln}
\end{figure}
%%%%%%%%%%%%%%%%%%%%%%%%%%%%%%%%%%%%%%%%%%%%%%%%%%%%%%%%%%%%%%%%%%%%
%The $SU(4)_C$ gauge symmetry is spontaneously broken to $SU(3)_C\otimes U(1)_{B-L}$ and $SU(2)_L\times SU(2)_R$ remain unbroken at the scale $M_I$. The matching 
%condition, therefore, is given as \cite{WEINBERG198051, Hall:1980kf,Bertolini:2009qj,Bertolini:2013vta, Chakrabortty:2019fov},
%%%%%%%%%%%%%%%%%%%%%%%%%%%%%%%%%%%%%%%%%%%%%%%%%%%%%%

During the second intermediate symmetry breaking at $M_{II}$, the diagonal generator of $SU(2)_R$ combines with $(B-L)/2$ to give the SM hypercharge $Y$. The 
GUT normalization for $B-L$ and $Y$ is $\sqrt{3/8}$ and $\sqrt{3/5}$ respectively. Therefore, the matching condition reads 
\begin{align}
\frac{1}{\alpha_{Y}(M_{II})} &= \frac{3}{5}\left(\frac{1}{\alpha_{2R}(M_{II})}-\frac{1}{6\pi}\right) + \frac{2}{5} \frac{1}{\alpha_{B-L}(M_{II})} - 
\frac{\lambda_Y (M_{II})}{12\pi} \ .
\end{align}
%%%%%%%%%%%%%%%%%%%%%%%%%%%%%%%
%%%%%%%%%%%%%%%%%%%%%%%%%%%%%%%%%%%%%%%%%%%%%%%%
 Fig.~\ref{fig:unific_soln} shows the unification solutions of the model with the simplified assumption that the heavy multiplets 
that decoupled during the successive symmetry breakings are degenerate with the 
respective breaking scale. In this case, the threshold corrections given in the Appendix vanish. We see that in this case, the proton
lifetime is only marginally within the reach of future experiments.
 
We next study the non-trivial effect of the threshold corrections assuming that the heavy multiplets are not degenerate with the 
respective breaking scales \cite{Langacker:1992rq, PhysRevD.47.R4830, PhysRevD.47.264, PhysRevD.49.3711, Parida:1995td,Dorsner:2005fq, Li:2009fq, Bertolini:2012im, Bertolini:2013vta, Babu:2015bna, Schwichtenberg:2018cka,Chakrabortty:2019fov,Ohlsson:2020rjc}. The threshold corrections in the Appendix will then have non-zero contributions. 
%%%%%%%%%%%%%%%%%%%%%%%%%%%%%%%%%%%%%%%%%%%%%%%%%%%%%%%%%%

We next consider an effective dimension-5 operator. The renormalizable gauge kinetic term in $SO(10)$ is
\begin{align}\label{eq:kin-lag}
\mathcal{L}_{\rm kin} = -\frac{1}{4C}{\rm Tr} (F^{\mu\nu}  F_{\mu \nu}) .
\end{align}
Here, $F^{\mu\nu} = F^{\mu\nu}_iT^i$ with $T^i$ being the generators of $SO(10)$ normalized as ${\rm Tr} (T^i  T^j) = C\delta^{ij}$.
The $210$-dimensional scalar $\chi_{210}$ comes from the symmetric product of two adjoint representations of $SO(10)$, i.e. 
$210\subset (45\times 45)_S$. We can write an effective dimension-5 operator given by
\begin{align}\label{eq:dim-5-oper}
\mathcal{L}_{\rm kin}^{\rm dim\text{-}5} = -\frac{\eta}{\Lambda} \Big[\frac{1}{4C}{\rm Tr} (F^{\mu\nu} \chi_{210} F_{\mu \nu})\Big ].
\end{align}
The cut-off scale $\Lambda$ in the Wilson coefficient $(\eta/\Lambda)$ of Eq.~(\ref{eq:dim-5-oper}) could be of the order of the 
reduced Planck scale $m_{\rm Pl}$. As $\chi_{210}$ acquires a non-zero VEV, the operator in Eq.~(\ref{eq:dim-5-oper}), in addition to the 
threshold corrections also modifies the unification boundary conditions \cite{Shafi:1983gz,Hill:1983xh,Hall:1992kq,Chakrabortty:2008zk,Chakrabortty:2009xm,Preda:2022izo}:
\begin{align}\label{eq:uni-bound-dim-5}
g_U^2=g_i^2(M_U)(1+\varepsilon \delta_i) ,
\end{align}  
   where $\varepsilon =\eta \left< \chi_{210} \right>/(2\Lambda) \sim \mathcal{O} (M_U/m_{\rm Pl})$, and the group-theoretic factors 
are $\delta_{2L} = 1/\sqrt{2}$, $\delta_{2R} = -1/\sqrt{2}$, and $\delta_{4C} = 0$. 
%%%%%%%%%%%%%%%%%%%%%%%%%%%%%%%%%%%%%%%%%%%%%%%%%%%%%%%%%
%%%%%%%%%%%%%%%%%%%%%%%%%%%%%%%%%%%%%%%%%%%
\begin{figure}[h!]
\begin{center}
\subfloat[{$m_S/M_i\in [1/2,2]$ ($i=I, \ II$)}]{\includegraphics[width=0.48\textwidth]{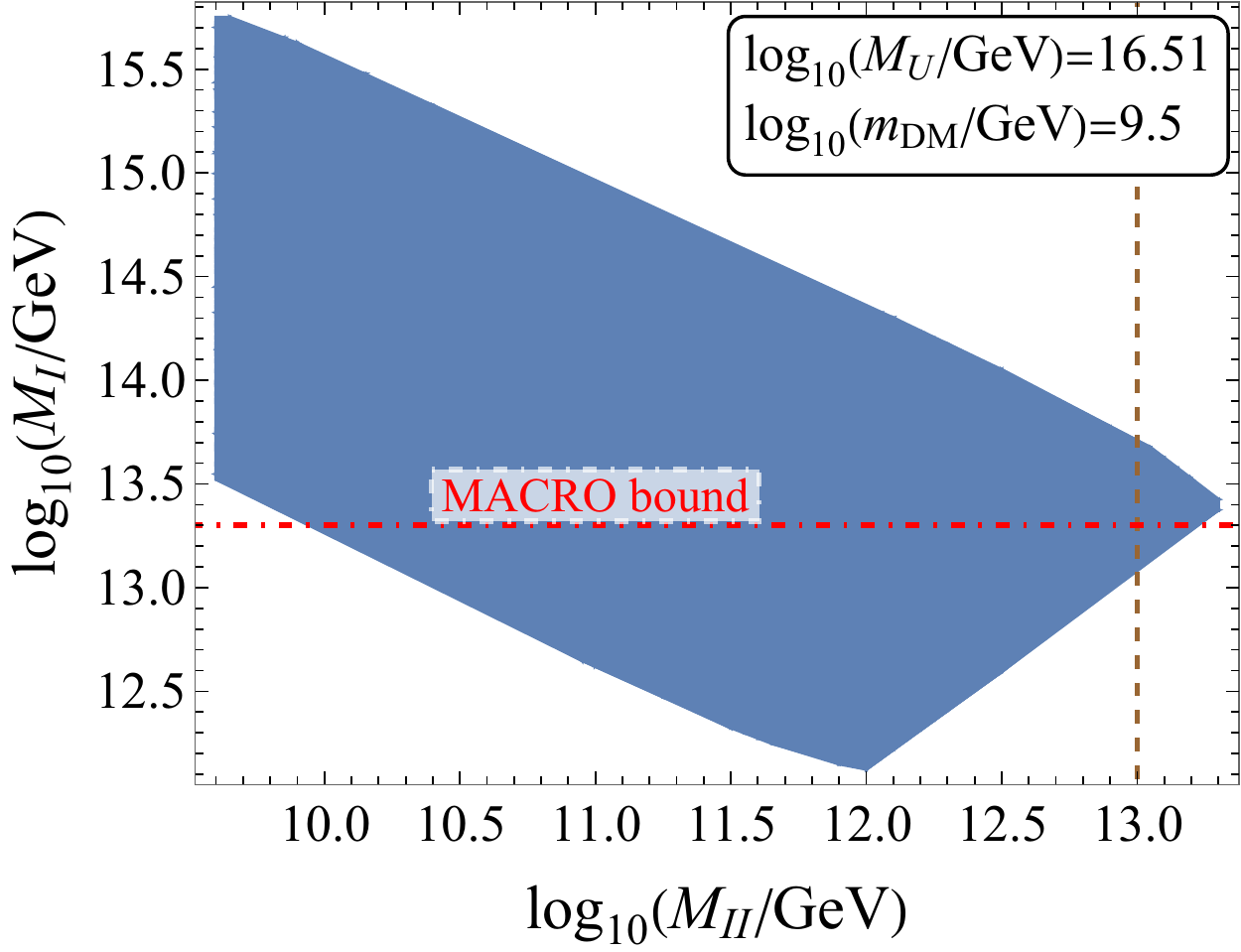}} \hspace{4mm}
\subfloat[{$m_S/M_i\in [1/5,5]$}]{\includegraphics[width=0.47\textwidth]{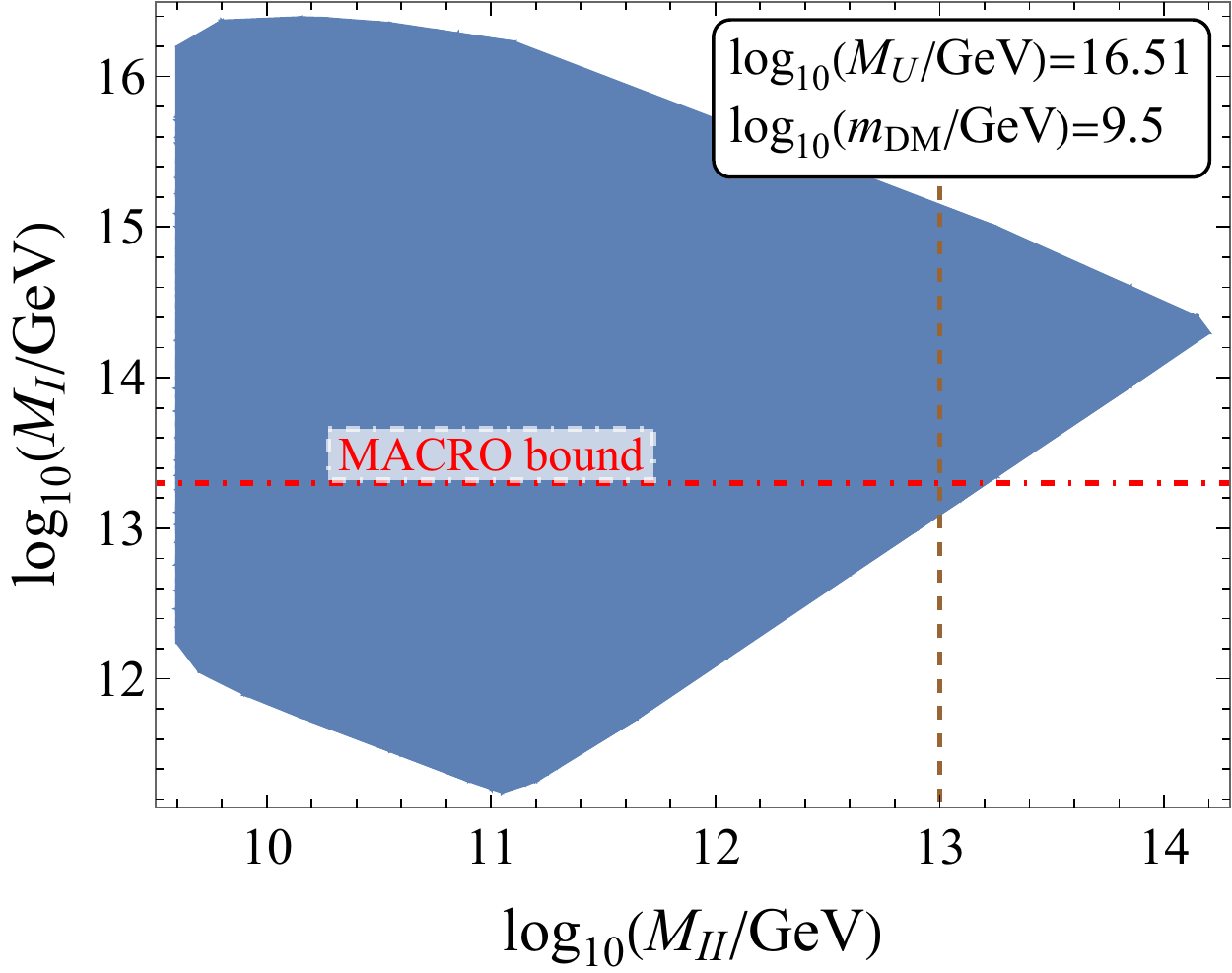}} \\
\subfloat[{$m_S/M_i\in [1/2,2]$ ($i=I, \ II$)}]{\includegraphics[width=0.48\textwidth]{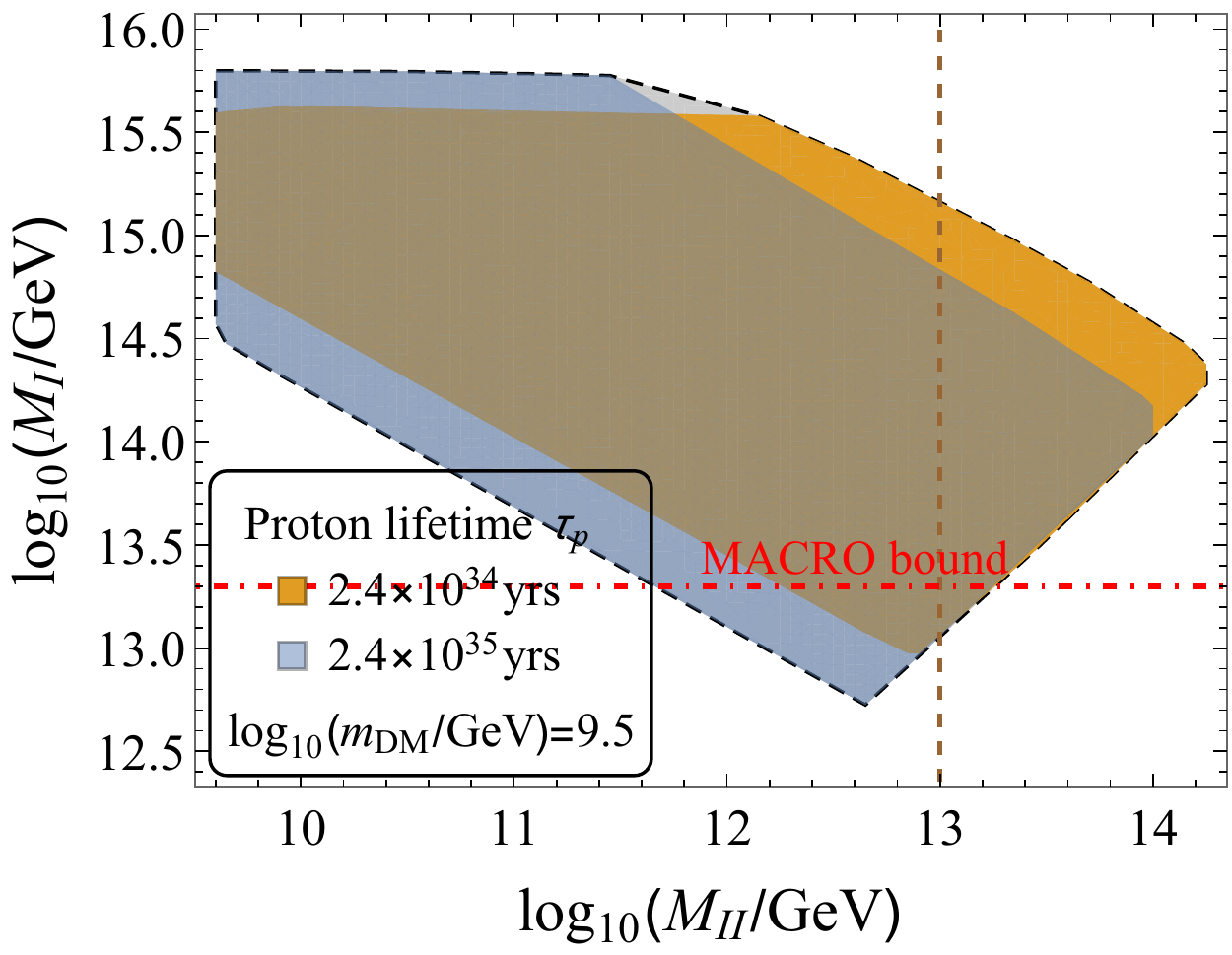}} \hspace{4mm}
\subfloat[{$m_S/M_i\in [1/5,5]$}]{\includegraphics[width=0.47\textwidth]{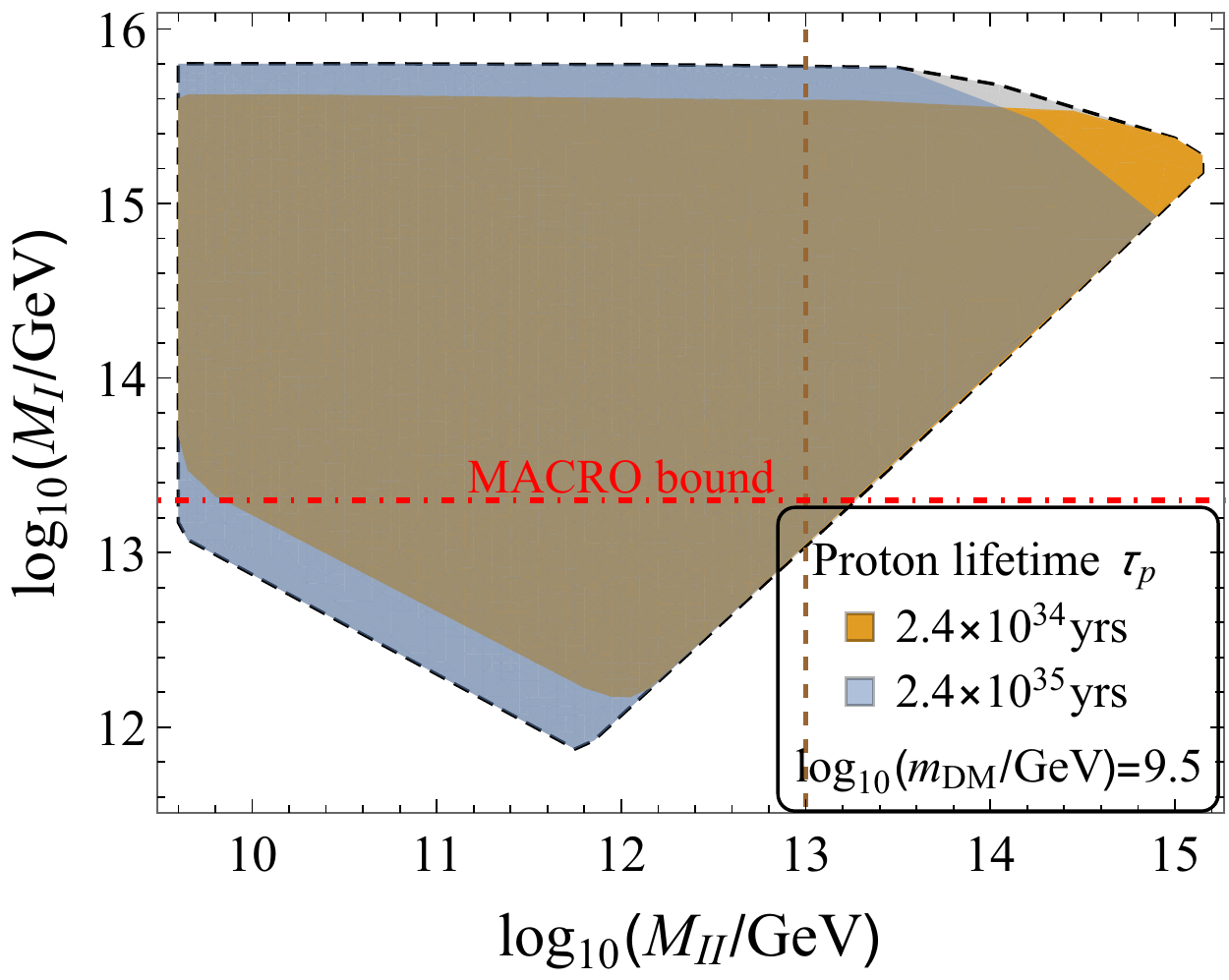}}
\end{center}
\caption{Plot for the intermediate scale $\log_{10} (M_I/\mathrm{GeV})$ versus $\log_{10} (M_{II}/\mathrm{GeV})$ with threshold corrections 
and dimension-5 operator included. In the upper panels, the unification scale is taken to be $\log_{10} (M_U/\mathrm{GeV}) 
= 16.51$ and the fermionic DM mass $\log_{10}(m_{\rm DM}/\mathrm{GeV}) = 9.5$. The MACRO bound \cite{Ambrosio:2002qq} on the flux of the 
$SU(4)_C\times SU(2)_L\times SU(2)_R$ monopoles associated with the scale $M_{I}$ is also depicted (see Section~\ref{sec:top}). The vertical 
dashed line corresponds to $M_{II}\sim 10^{13}~{\rm GeV}$ which is suitable 
for leptogenesis (see Section~\ref{sec:dm_lepto}). In the lower panels, we display the unification solutions with proton decay 
satisfying the Super-Kamiokande bound \cite{Super-Kamiokande:2020wjk} and observable in the Hyper-Kamiokande experiment 
\cite{Dealtry:2019ldr}.}\label{fig:unific_soln_thr_cor}
\end{figure}
%%%%%%%%%%%%%%%%%%%%%%%%%%%%%%%%%%%%%%%%%%%%%%%%%%%%%%%%%%%

The upper panels of Fig.~\ref{fig:unific_soln_thr_cor} show the solution region for successful inflation (see 
Section~\ref{sec:inflation}) with $\log_{10}(m_{\rm DM}/{\rm GeV})=9.5$ and $\log_{10}(M_U/{\rm  GeV})=16.51$ after 
including the threshold corrections in Eqs.~(\ref{eq:thr_cor_mU}), (\ref{eq:thr_cor_mI}), and (\ref{eq:thr_cor_mII}) and the effect of 
the dimension-5 operator in Eq.~(\ref{eq:dim-5-oper}). We assume that the heavy scalars have masses within $m_S/M_i\in [1/2,2]$ in panel 
(a) and $m_S/M_i\in [1/5,5]$ in panel (b) with $i=I,  II,  U$. The heavy multiplets in the daughter gauge symmetry which 
decouple are shown in Table~\ref{tab:RGE_multiplets}. We can have unification solutions with a breaking scale $M_{II}\sim 10^{13}$ GeV 
(see Fig.~\ref{fig:unific_soln_thr_cor}), which is suitable for a successful leptogenesis as we will see in Section~\ref{sec:dm_lepto}. 
The flux of the intermediate mass monopoles associated with the scale $M_I$ (see Section~\ref{sec:top}) can be measurable 
\cite{Lazarides:1980cc}. In the lower panels of Fig.~\ref{fig:unific_soln_thr_cor}, we display the unification solutions 
that yield proton decay lifetimes compatible with the Super-Kamiokande \cite{Super-Kamiokande:2020wjk} bound and which could be observed 
in the Hyper-Kamiokande experiment \cite{Dealtry:2019ldr}. The unification scale $M_U$ in this case ranges between about
$4.5\times 10^{15}~{\rm GeV}$ and $7.5\times 10^{15}~{\rm GeV}$. Successful inflation (see Section~\ref{sec:inflation}) will then 
require large values of the relevant coupling constant.

 %%%%%%%%%%%%%%%%%%%%%%%%%%%%%%%%%%%%%%%%%%%%%%%%%%%%%%%%%%%%%%%%%%
\section{Inflation with Coleman-Weinberg Potential}
\label{sec:inflation}
 We consider inflation driven by the Coleman-Weinberg potential of a real GUT-singlet inflaton field $\phi$ 
\cite{Shafi:1983bd,Lazarides:1984pq,Shafi:2006cs,Rehman:2008qs,Senoguz:2015lba}
\begin{equation}\label{CW-potential}
V(\phi)= A\phi^4\left[ \log\left(\frac{\phi}{M}\right) - \frac{1}{4}\right] +V_0,
\end{equation}
where $V_0=AM^4/4$ and the potential is minimized at $\phi=M$ with $V(\phi=M)=0$. Let us note in passing that 
a non-minimal coupling of the inflaton to gravity predicts smaller values for the tensor-to-scalar ratio $r$ 
\cite{Bostan:2018evz,Bostan:2019uvv,Bostan:2019fvk,Maji:2022jzu}, which are preferred by the recent data \cite{BICEP:2021xfz} at $95\%$ confidence level.
%%%%%%%%%%%%%%%%%%%%%%%%%%%%%%%%%%%%%%%%%%%%%%%%%%%%%%%%%%%%%%%%
\begin{figure}[htbp!]
\centering
\includegraphics[scale=0.6]{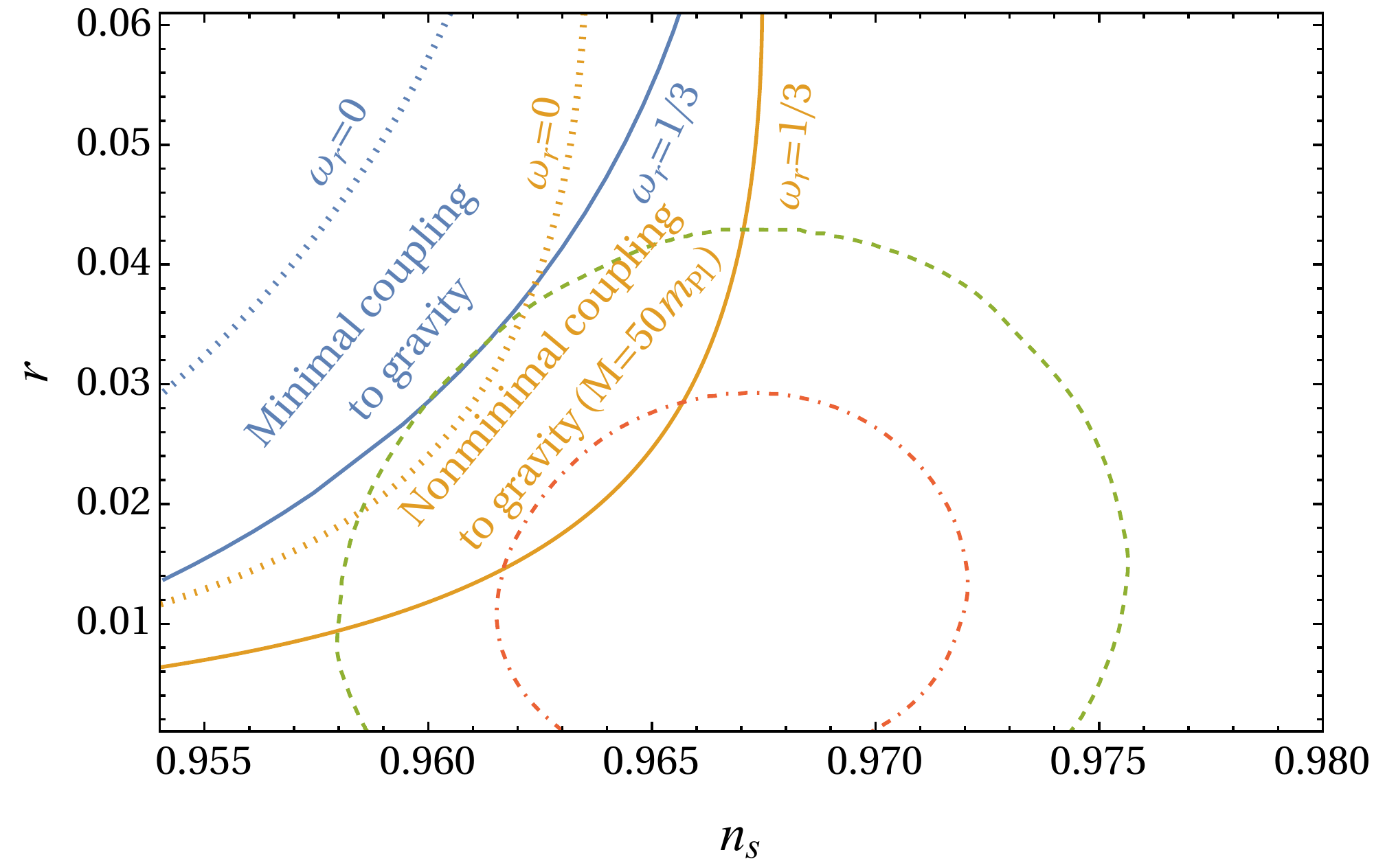}
\caption{Predictions for $n_{\rm s}$ and $r$ for inflation driven by the Coleman-Weinberg potential of a real GUT singlet with minimal and 
nonminimal coupling to gravity. We take two values for $\omega_r$, the effective equation-of-state
parameter from the end of inflation until reheating. The dot-dashed and dashed contours are the 68\% and 95\% confidence level contours of
Planck TT, TE, EE + lowE + lensing + BK18 + BAO \cite{BICEP:2021xfz}.}\label{Fig:min-nomin}
\end{figure} 
%%%%%%%%%%%%%%%%%%%%%%%%%%%%%%%%%%%%%%%%%%%%%%%%%%%%%%%%%%%%%%%%
 Fig.~\ref{Fig:min-nomin} compares the predictions for $n_{\rm s}$ and $r$ when the inflaton $\phi$ has minimal and nonminimal 
coupling to gravity. The 68\% and 95\% confidence level contours of
Planck TT, TE, EE + lowE + lensing + BK18 + BAO \cite{BICEP:2021xfz} are also depicted. For nonminimal coupling to gravity, we have 
taken $M=50m_{\rm Pl}$ (for details see Ref.~\cite{Maji:2022jzu}). The interaction potential that induces the 
VEVs of the scalars 
at different symmetry-breaking scales is given by
\begin{align}\label{eq:pot}
V(\phi,\chi_D) = - \frac{1}{2}\beta_D^2\phi^2\chi_D^2 + \frac{\lambda_D}{4}\chi_D^4, 
\end{align}
where the symmetry-breaking scalars $\chi_D$ are canonically normalized real scalar fields with $D$ being the dimensionality of the representation to which $\chi_D$ belongs. The final VEVs are given by
\begin{align}
\left<\chi_D\right> = (\beta_D/\sqrt{\lambda_D}) M .
\end{align}
The interaction potential in Eq.~(\ref{eq:pot}) gives rise to the coefficient $A=\beta_D^4\, D/16\pi^2$ \cite{Lazarides:2019xai} 
in Eq.~(\ref{CW-potential}). This coefficient is dominated by the coupling $\beta_{210}$ of the scalar $\chi_{210}$ which acquires 
a non-zero VEV at $M_U$. Therefore, we can write 
\begin{align}\label{eq:mU_2}
M_U\equiv\left<\chi_{210}\right> = \sqrt{\frac{8\pi}{\lambda_{210}}}\left(\frac{V_0}{210}\right)^{1/4}.
\end{align}

The effective mass squared of $\chi_D$ at the completion of the corresponding phase transition is given by
\begin{align}
m_{\rm eff}^2 = 2 (\beta_D^2\phi^2 - \sigma_{\chi_D} T_H^2) ,
\end{align}
where $T_H=H/2\pi$ is the Hawking temperature during inflation with $H$ being the Hubble parameter and $\sigma_{\chi_D} \sim 1$. 
%Difference of potential between local maximum and global minima:
%\begin{align}
%\Delta V = \frac{(\beta^2\phi^2-\sigma_\chi T_H^2)^2}{4a} = \frac{m_{\rm eff}^4}{16a} \ .
%\end{align}
The completion of the phase transition is governed by the Ginzburg criterion $\xi^3 \Delta V > T_H$ \cite{ginzburg}. Here, 
$\xi \sim \mathrm{min} [ H^{-1}, m_{\rm eff}^{-1} ]$ is the correlation length and $\Delta V=m_{\rm eff}^4/(16\lambda_D)$ is 
the difference between the potential at  $\chi_D = 0$ and $\left< \chi_D \right>$. The phase transitions occur when
\begin{align}\label{eq:phase_tran}
\beta_D \phi = \begin{cases}
 \sqrt{\left(128 \lambda_D^2 + \sigma_{\chi_D}\right)} \ \frac{H}{2\pi} & \mathrm{for} \ \ m_{\rm eff}^{-1} \lesssim H^{-1} , \\
 \sqrt{\left(4\pi\sqrt{2\pi \lambda_D} + \sigma_{\chi_D}\right)} \ \frac{H}{2\pi} & \mathrm{for} \ \ m_{\rm eff}^{-1} \gtrsim H^{-1} .
 \end{cases}
\end{align}

  Successful inflation compatible with the Planck 2018 data \cite{Planck:2018jri} occurs for a typical choice $V_0^{1/4} = 
1.75\times 10^{16}$ GeV, $A = 1.43\times 10^{-14}$, and $M = 7.17\times 10^{19}$ GeV. The corresponding unification scale for 
$\lambda_{210} = 1/2$ is $M_U=3.26\times 10^{16}$ GeV \cite{Chakrabortty:2020otp} as one can deduce 
from Eq.~(\ref{eq:mU_2}). For $M_U=7.5\times 10^{15}$ GeV, which may lead to proton lifetime measurable in 
Hyper-Kamiokande \cite{Dealtry:2019ldr}, the required value of $\lambda_{210}$ is 9.44. 
%%%%%%%%%%%%%%%%%%%%%%%%%%%%%%%%%%%%%%%%%%%%%%%%%%%%%%%%%%%%%%%%%%
\section{Monopoles, Strings, and Gravitational Waves}
\label{sec:top}
The symmetry breaking scheme in Eq.~(\ref{chain}) yields superheavy GUT monopoles of mass $\sim 10M_U$ \cite{Lazarides:2019xai,Lazarides:1980cc}, 
intermediate scale monopoles of mass $\sim 10 M_I$ \cite{Lazarides:1980cc}, and topologically stable cosmic strings associated with the 
scale $M_{II}$ \cite{Kibble:1982ae}. 
%The spontaneous symmetry breaking at the unification scale generates GUT-scale monopoles \cite{Lazarides:1980cc}, which become Dirac monopoles after the 
%electroweak symmetry breaking \cite{Lazarides:2019xai}. The spontaneous breaking of $SU(4)_C$ to $SU(3)_C\times U(1)_{B-L}$ forms intermediate mass 
%monopoles which are topologically stable and distinct from the GUT-scale monopoles. After the electroweak symmetry breaking, they develop into Schwinger 
%monopoles with two units of Dirac magnetic charge \cite{Lazarides:1980cc}. The next breaking $SU(2)_R\times U(1)_{B-L}\to U(1)_Y\times Z_2$ at $M_{II}$ 
%generates local cosmic strings which are topologically stable \cite{Kibble:1982ae}. %{\color{red}and 't Hooft-Polyakov-type monopoles 
%\cite{tHooft:1974kcl,Polyakov:1974ek} (Could you please explain the fate of such monopoles?)}. 
The dimensionless tension of the strings, in the Bogomol'nyi limit of the Abelian Higgs model, is given by
\begin{align}
G\mu = \frac{1}{8}\left(\frac{\left<\phi_{126}\right>}{m_{\rm Pl}}\right)^2 ,
\end{align}
where $G$ is Newton's gravitational constant and we have $\left<\phi_{126}\right>=M_{II}$.
These strings intercommute generating string loops that decay via the radiation of gravitational waves \cite{Vachaspati:1984gt}. The 
production of gravitational waves from cosmic strings and related hybrid defects has been discussed in literature including 
Refs.~\cite{Martin:1996ea,Martin:1996cp,Vilenkin:2000jqa,Leblond:2009fq,Sousa:2013aaa,Cui:2017ufi,Cui:2018rwi,Guedes:2018afo,
Gouttenoire:2019kij,Buchmuller:2019gfy,King:2020hyd,Ellis:2020ena,Buchmuller:2020lbh,King:2021gmj,
Buchmuller:2021dtt,Buchmuller:2021mbb,Masoud:2021prr,Dunsky:2021tih,Chun:2021brv,Afzal:2022vjx,Ahmed:2022rwy,Lazarides:2022jgr,Fu:2022lrn}.

The gravitational wave background at a frequency $f$ is given by \cite{Olmez:2010bi,Auclair:2019wcv,Cui:2019kkd,LIGOScientific:2021nrg}
\begin{align}\label{eq:GWs-Omega-cusps}
\Omega_{\rm GW}(f) = \frac{4\pi^2}{3H_0^2}f^3\int_{z_*}^{z(t_F)}dz \int dl \, h^2(f,l,z)\frac{d^2R}{dz \, dl} \ ,
\end{align}
where the waveform assuming cusp domination is given by
\begin{align}
h(f,l,z) = g_{1c}\frac{G\mu \, l^{2/3}}{(1+z)^{1/3}r(z)}f^{-4/3},
\end{align}
with $g_{1c}\simeq 0.85$ \cite{LIGOScientific:2021nrg}, $z$ being the redshift, and $l$ the loop length. 
The time $t_F$ corresponds to the onset of loop formation, and the lower limit $z_*$ in Eq.~(\ref{eq:GWs-Omega-cusps}) 
excludes the infrequent bursts from the stochastic background such that \cite{Cui:2019kkd}
\begin{align}
\int_0^{z_*} dz  \int dl \,\frac{d^2R}{dz dl} = f .
\end{align}
The proper distance $r$ is given by
\begin{align}
r(z)=\int_0^z\frac{dz'}{H(z')} ,
\end{align}
where $H$ is the Hubble parameter with its present value denoted by $H_0$.
The burst rate per unit space-time volume is given by
\begin{align}
\frac{d^2R}{dz \, dl} = N_c H_0^{-3}\phi_V(z) \frac{2n(l,t(z))}{l(1+z)}\left( \frac{\theta_m(f,l,z)}{2}\right)^2\Theta(1-\theta_m),
\end{align}
where we have set $N_c=2.13$ as in Ref.~\cite{Cui:2019kkd}, the beam opening angle
\begin{align}
\theta_m(f,l,z) = \left[\frac{\sqrt{3}}{4}(1+z)fl\right]^{-1/3} ,
\end{align}
and
\begin{align}
\phi_V(z) = \frac{4\pi H_0^3r^2}{(1+z)^3H(z)} \ .
\end{align} 
In the radiation dominated universe the loop distribution function $n(l,t)$ at the time of gravitational wave emission is given by 
\cite{Blanco-Pillado:2013qja,Blanco-Pillado:2017oxo}
\begin{align}\label{eq:n-loop-rad}
n_r(l,t) = \frac{0.18}{t^{3/2}(l+\Gamma G\mu t)^{5/2}}\Theta(0.18t-l) \ .
\end{align}
%%%%%%%%%%%%%%%%%%%%%%%%%%%%%%%%%%%%%%%%
\begin{figure}[h!]
\centering
\includegraphics[width=0.7\linewidth]{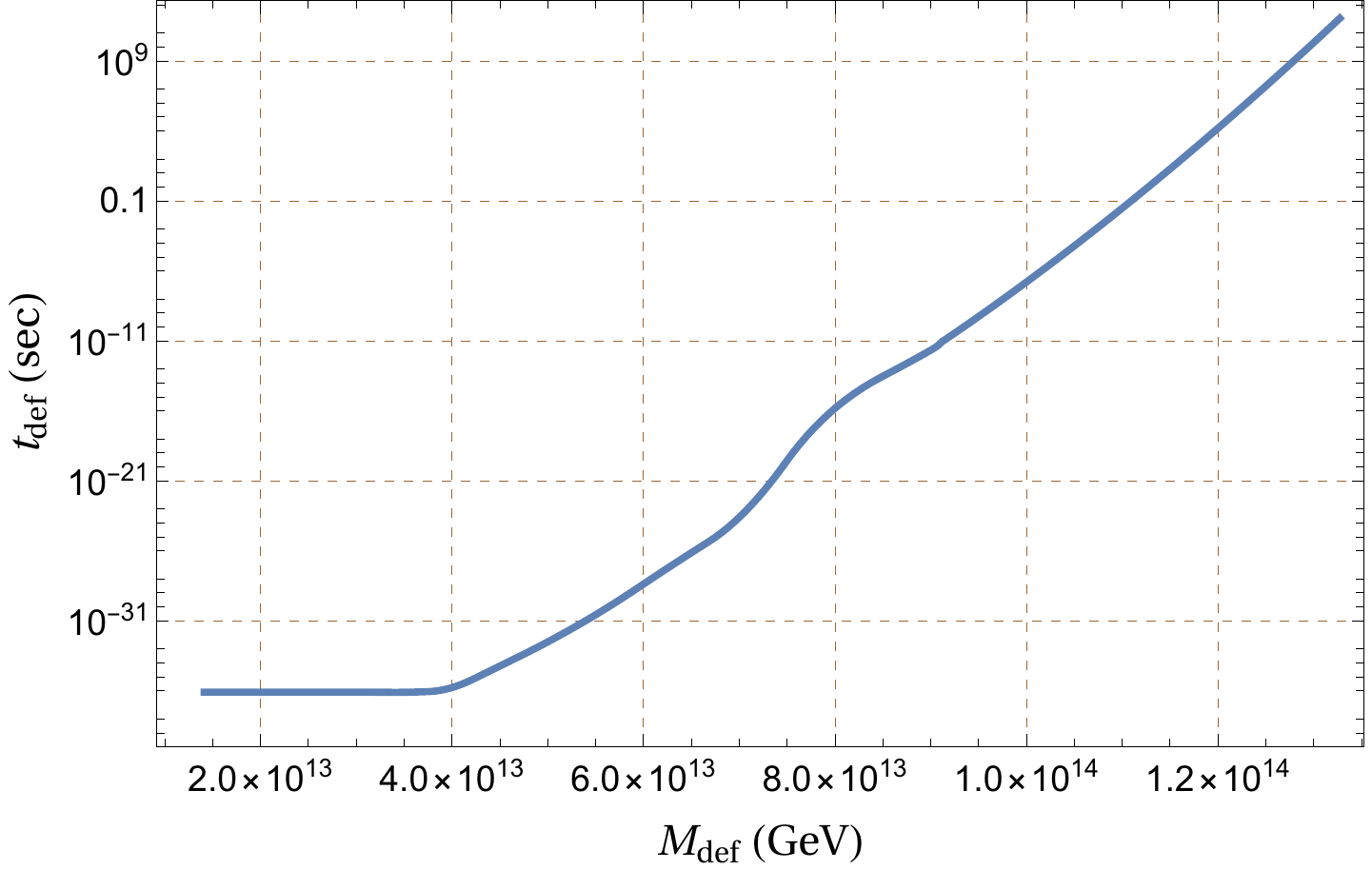}
\caption{Horizon reentry time $t_{\rm def}$ of topological defects (strings or monopoles) formed during inflation as a function of the 
symmetry breaking scale $M_{\rm def}$.}\label{plt:t_mI}
\end{figure}
%%%%%%%%%%%%%%%%%%%%%%%%%%%%%%%%%%%%%%%%%%%%%%%%
%%%%%%%%%%%%%%%%%%%%%%%%%%%%%%%%%%%%%%%%%%%%
\begin{figure}[htbp!]
\centering
\subfloat[][The gravitational wave spectra for strings present or reentering the horizon at $t_F\lesssim 10^{-20}$ sec. We have safely taken 
$t_F = 10^{-20}$ sec to compute the spectra. The numbers are the values of $\log_{10}(M_{II}/{\rm GeV})$ for the respective spectrum.]
{\includegraphics[width=0.7\linewidth]{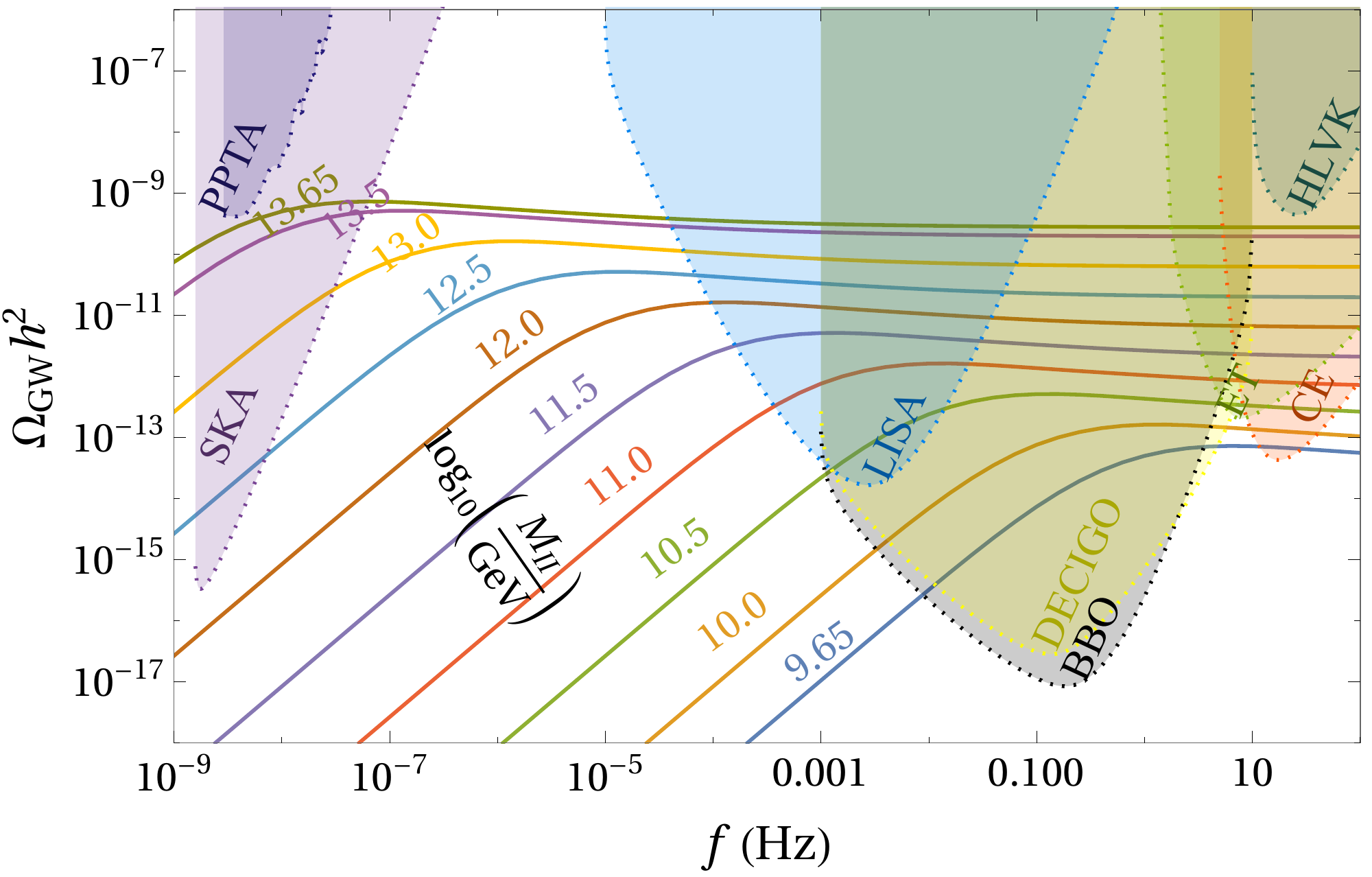}\label{fig:GWs-a}}\\
\subfloat[][The gravitational wave spectra for strings reentering the horizon at $t_F\gtrsim 10^{-20}$ sec. The numbers in the parentheses 
are the values of $\log_{10}(M_{II}/{\rm GeV})$ and $\log_{10}(t_F/{\rm sec})$ for the respective spectrum.]
{\includegraphics[width=0.7\linewidth]{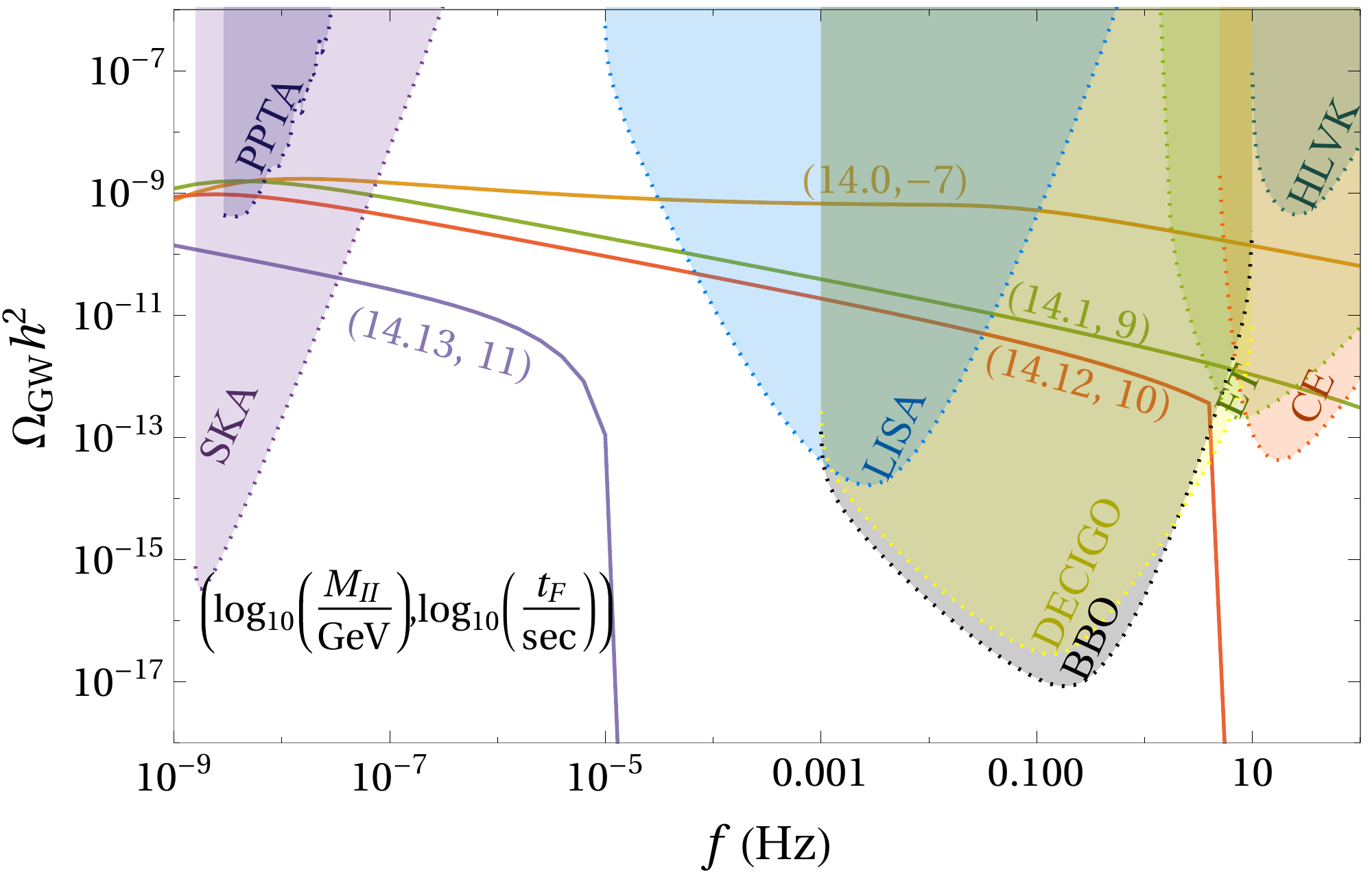}\label{fig:GWs-b}}
\caption{Gravitational wave spectra from cosmic strings generated during the spontaneous symmetry breaking at the scale $M_{II}$. The 
strings formed at the breaking scales $M_{II}\in [10^{9.65},10^{13.65}]$ GeV appear in the horizon at a very early cosmic time 
$t_F\lesssim 10^{-20}$ sec. The resulting gravitational wave background satisfies the present PPTA exclusion limit and can be probed by 
some of the proposed experiments. The strings formed during inflation around a breaking scale $1.3\times 10^{14}$ GeV reenter the horizon 
at $t_F\simeq 10^{11}$ sec and the gravitational wave spectrum from them satisfies the PPTA bound and can be probed by the SKA experiment. 
The sensitivity curves \cite{Thrane:2013oya, Schmitz:2020syl} for PPTA \cite{Shannon:2015ect} and various proposed experiments, including 
SKA \cite{5136190, Janssen:2014dka}, CE \cite{PhysRevLett.118.151105}, ET \cite{Mentasti:2020yyd}, 
LISA \cite{Bartolo:2016ami, amaroseoane2017laser}, DECIGO \cite{Sato_2017}, BBO \cite{Crowder:2005nr, Corbin:2005ny}, 
HLVK \cite{KAGRA:2013rdx}, are also shown in the plots.}\label{fig:GWs}
\end{figure}
%%%%%%%%%%%%%%%%%%%%%%%%%%%%%%%%%%%%%%%
In the matter-dominated Universe, there are two contributions. For loops that are remnants from the radiation era, 
\begin{align}\label{eq:n-loop-radmat}
n_{rm}(l,t) = \frac{0.18t_{eq}^{1/2}}{t^2(l+\Gamma G\mu t)^2}\Theta(0.18t_{eq}-l-\Gamma G\mu(t-t_{eq})) \ ,
\end{align}
where $t_{eq}$ is the equidensity time. For loops which are produced during the matter-dominated era, 
\begin{align}\label{eq:n-loop-mat}
n_{m}(l,t) = \frac{0.27-0.45(l/t)^{0.31}}{t^2(l+\Gamma G\mu t)^2}\Theta(0.18t-l)\Theta(l+\Gamma G\mu(t-t_{eq})-0.18t_{eq}) \ .
\end{align} 
We have taken the integration limits on $l$ to be from $0$ to $2t$ ($3t$) for radiation (matter) domination. However, the various 
Heaviside $\Theta$ functions will control the upper and lower limits during numerical evaluations. 

Fig.~\ref{plt:t_mI} shows the horizon reentry time of topological defects such as strings or monopoles that are generated at the symmetry breaking 
scale $M_{II}$ or $M_I$ (call it $M_{\rm def}$) during inflation. Following the analysis of  Ref.~\cite{Lazarides:2022jgr}, we find that in our 
case inflation ends at a cosmic time $7.8\times 10^{-37}$ sec. Assuming $\lambda_D\sim g^2$ and using Eq.~(\ref{eq:phase_tran}), we see that a 
spontaneous symmetry breaking occurs during inflation only if the breaking scale $M_{\rm def}\gtrsim 1.5\times 10^{13}$ GeV (see 
Refs.~\cite{Chakrabortty:2020otp,Lazarides:2021uxv} for more details). The gravitational wave spectra from the strings formed at the breaking 
scales $M_{II}\in [10^{9.65},10^{13.65}]$ GeV are shown in Fig.~\ref{fig:GWs-a}. The present PPTA upper bound \cite{Shannon:2015ect} is saturated 
for strings that are formed during inflation at the breaking scale $4.5\times 10^{13}$ GeV. In this case the strings reenter the horizon at a 
very early time $t_F<<t_{eq}\simeq 1.48\times 10^{12}$ sec as we can see from Fig.~\ref{plt:t_mI}.
%%%%%%%%%%%%%%%%%%%%%%%%%%%%%%%%%%%%%%%%%%%%%%%%%%%%%%%%%%%%%%%%
 Actually, in all cases depicted in 
Fig.~\ref{fig:GWs-a} the strings are generated either during inflation and reenter the horizon at a very early time $t_F<<t_{eq}$, or during 
inflaton oscillations following the end of inflation at $t_F\lesssim 10^{-20}$ sec. Consequently, the gravitational wave spectrum remains 
unaffected by inflation in the nHz to kHz regime. Note that the phase transitions after the end of inflation are governed by the radiation 
temperature from the inflaton decay since this temperature soon outstrips the Hawking temperature (see Ref.~\cite{Chakrabortty:2020otp}). The 
radiation temperature approaches the reheat temperature ($\sim 2\times 10^6$ GeV) at the reheat time $t_r\simeq 6.8 \times 10^{-18}$ sec. 
%Therefore, we have safely chosen $t_F = 10^{-20}$ sec to compute the gravitational wave spectra for $10^9~\mathrm{GeV} <M_{II}<10^{13}~\mathrm{GeV}$. 
We have checked that the contribution to the gravitational wave spectrum from the loops generated during inflaton oscillations is quite negligible 
in the frequency range under consideration and we therefore set $t_F = 10^{-20}$ sec (see also Ref.~\cite{Lazarides:2022jgr}).  
Various proposed experiments, including the SKA \cite{5136190, Janssen:2014dka}, CE \cite{PhysRevLett.118.151105}, ET \cite{Mentasti:2020yyd}, 
LISA \cite{Bartolo:2016ami, amaroseoane2017laser}, DECIGO \cite{Sato_2017}, and BBO \cite{Crowder:2005nr, Corbin:2005ny} experiments, can 
observe this stochastic gravitational wave background. 
%Soon after inflation the radiation temperature from the inflaton decay outstrips the Hawking temperature and governs the phase transitions 
%after the inflation. The radiation temperature approaches the reheat temperature ($\sim 2\times 10^6$ GeV) at the reheat time 
%$t_r\simeq 6.8 \times 10^{-18}$ sec. Therefore, we have safely chosen $t_F = 10^{-20}$ sec to compute the gravitational wave 
%spectra for $10^9~\mathrm{GeV} <M_{II}<10^{13}~\mathrm{GeV}$. 
Fig.~\ref{fig:GWs-b} shows the gravitational wave background for strings formed during inflation at the scales $M_{II}\gtrsim 4.5\times 10^{13}$ GeV. 
We see that the gravitational wave background from strings formed around a breaking scale $1.3\times 10^{14}$ GeV that reenter the horizon at 
$t_F\simeq 10^{11}$ sec is the only one satisfying the PPTA bound. This can be probed by the SKA experiment.

%%%%%%%%%%%%%%%%%%%%%%%%%%%%%%%%%%%%%%%%%%%%%%%%%%%%%%%%%%%%%%%%%
%%%%%%%%%%%%%%%%%%%%%%%%%%%%%%%%%%%%%%%%%%%%%%%%%%%%%%%%
\section{Non-thermal Dark Matter and Leptogenesis }
\label{sec:dm_lepto}
The observed DM relic density can be realized in our setup non-thermally from the inflaton's decay and axion. 
Moreover, the baryon asymmetry in the Universe can be generated via the right-handed neutrinos (RHNs) from the inflaton decay. 
The RHNs produce a primordial 
lepton asymmetry which is partially converted into the observed baryon asymmetry with the help of electroweak sphaleron effects. 
In practice, one needs to solve a set of coupled Boltzmann equations to investigate how these species evolve in the Universe. Following the 
prescription of Refs.~\cite{Hahn-Woernle:2008tsk,Barman:2021tgt,Barman:2021ost}, we construct the Boltzmann equations required 
for the analysis of the evolution of the different species involved in our setup. 

 Before delving into the details of Boltzmann equations, we would like to comment on the nature of the intermediate scale 
fermionic DM. The $\psi_{10}^{\alpha}~(\alpha = 1,2)$ offer two neutral Dirac fermions, the lightest 
of which can play the role of DM stabilized by the unbroken $Z_2$ symmetry of $SO(10)$. The heavier color triplets and anti-triplets 
of the two 10-plets can decay into the $SU(2)_L$ doublet in the same 10-plet and SM particles via mediation of heavy lepto-quark gauge 
bosons \cite{Lazarides:2020frf,Okada:2022yvq}. Apart from these, the charged members belonging to the $SU(2)_L$ doublets of these 
fermionic 10-plets can decay to the neutral components of the same doublet and SM particle -- see Ref.~\cite{Cirelli:2005uq} for details. 
It is interesting to point out that at the tree level, the two neutral fermions remain mass degenerate but a tiny non-zero mass splitting 
can be generated between them at the loop level. We refrain from going into the details of the calculation of this splitting and refer 
the reader to Refs.~\cite{Kadastik:2009dj,Mambrini:2015vna,Boucenna:2015sdg, Ferrari:2018rey,Okada:2019bqa} for more details.

\subsection{Boltzmann Equations}
In this section, we provide a set of coupled Boltzmann equations for the time evolution of a system comprised of an unstable massive particle 
$\phi$ (inflaton) with mass $m_{\phi}$, unstable RHNs ($N_i$) with mass $M_i$, radiation (\textit{R}), lepton number asymmetries generated 
in the visible sector 
($n_{L_i}$) as a result of the decay of the RHNs $N_i$, and a stable massive fermion $\psi_{\rm DM}$ (the lightest neutral component of 
$\psi_{10}$) with mass $m_{\rm DM}$ which contributes to DM. In this scenario, it is assumed that $\phi$ decays predominantly to RHNs but 
also to stable DM fermions. 

\beqs
\bea
\frac{d\rho_{\phi}}{dt}&=&-3H\rho_{\phi}-\G_{\phi}\rho_{\phi},  \\
\frac{d\rho_{N_i}}{dt}&=&-3H\rho_{N_i}+\G_{\phi\to N_iN_i}\rho_{\phi}-\G_{N_i}\rho_{N_i}, \\
\frac{d\rho_{\rm R}}{dt}&=&-4H\rho_{\rm R}+\sum_{i=1,2,3}\G_{N_i}\rho_{N_i}+\bra\sigma v\ket 2\bra E_{\psi_{\rm DM}}\ket n_{\psi_{\rm DM}}^2,\\
\frac{dn_{L_i}}{dt}&=&-3Hn_{L_i}+\epsilon_i\G_{N_i}n_{N_i}, \\
\frac{dn_{\psi_{\rm DM}}}{dt}&=&-3Hn_{\psi_{\rm DM}}+\G_{\phi\to \psi_{\rm DM}\psi_{\rm DM}}\frac{\rho_{\phi}}{m_{\phi}}-\bra\sigma v\ket n_{\psi_{\rm DM}}^2.   
\eea 
\label{be}
\eeqs

Here $\rho_i$ and $n_i$ represent the energy and the number densities of the particles under consideration, 
$\G_{\phi}=\G_{\phi\to N_iN_i}+\G_{\phi\to \psi_{\rm DM}\psi_{\rm DM}}$ 
denotes the total decay width of the inflaton, where 
\begin{equation}
\G_{\phi\to N_iN_i}=\frac{1}{16\pi}\bigg(\frac{M_i }{M}\frac{m_{126}^2}{ m_{\phi}^2}\bigg)^2m_{\phi} \quad {\rm and} \quad 
\G_{\phi\to \psi_{\rm DM}\psi_{\rm DM}}=\frac{10}{16\pi}\bigg(\frac{m_{\rm DM}}{M}\frac{m_{45}^2}{ m_{\phi}^2}\bigg)^2m_{\phi}
\end{equation} 
are the decay widths of the inflaton to RHNs and fermionic DM respectively (see Ref.~\cite{Lazarides:2020frf}) with $m_{126}$ 
and $m_{45}$ being the intermediate scale masses of $\phi_{126}$ and $\phi_{45}$. Also 
$\G_{N_i}=(1/8\pi)(y_{\nu}^{\dagger}y_{\nu})_{ii}M_i$ are the 
decay widths of the RHNs with $y_\nu$ being the neutrino Yukawa coupling matrix in the basis where the RHN masses are diagonal. 
The Hubble parameter is $H=\dot{a}/a$, 
where $a$ denotes the scale factor of the Universe and the overdot the time derivative. Moreover, $\bra\sigma v\ket$ 
corresponds to the thermal average of the annihilation 
cross section times the velocity of the DM fermions and we assumed that each DM fermion has energy $\bra E_{\psi_{\rm DM}}\ket$  
(with $\rho_{\psi_{\rm DM}}=\bra E_{\psi_{\rm DM}}\ket n_{\psi_{\rm DM}}$). Finally, the lepton asymmetry ($\epsilon_i$) generated by the decay 
of a RHN $N_i$ is given as~\cite{Covi:1996wh}
\beq
\label{cpasym}
\e_i=\frac{1}{8\pi} \frac{\text{Im}[(y_{\nu}^\dagger y_{\nu})^2_{ij}]}{(y_{\nu}^\dagger y_{\nu})_{ii}}\mathcal{F}\left(\frac{M_j^2}{M_i^2}\right),
\eeq
where $\mathcal{F}(x)=\sqrt{x}\bigg[1+\frac{1}{1-x}+(1+x)\ln(\frac{x}{1+x})\bigg]$.

In Eq.~(\ref{be}a) we show the evolution of the inflaton energy density. The first term in the right-hand side (RHS) of this equation is responsible for the 
dilution of the inflaton energy density due to the expansion of the Universe, and the second term accounts for the reduction of this energy density 
due to the inflaton decay.  Eq.~(\ref{be}b) describes the evolution of the energy density of the RHNs. Here the first term in the RHS again shows the 
dilution effect due to the expansion of the Universe, the second term is due to the production of the RHN from the decay of the inflaton (hence the 
positive sign), and the third term accounts for the depletion (negative sign) of the RHN energy density due to its decay into SM particles. 
Next, we provide the evolution of the radiation energy density in Eq.~(\ref{be}c). Here, the first term in the RHS represents the dilution of the 
radiation energy density due to the expansion and comes with a factor of 4 rather than 3 because the radiation energy density scales as $\rho_R \propto 
a^{-4}$. The second 
term depicts the production of radiation from the decay of the RHNs, and the third term shows the contribution coming from the annihilation of DM 
fermions into radiation with the factor of 2$\bra E_{\psi_{\rm DM}}\ket$ being the average energy released in a pair annihilation of DM fermions. In 
Eq.~(\ref{be}d) we show the evolution of the lepton number asymmetry $n_{L_i}$ generated in the visible sector due to the decay of the RHN $N_i$ to 
SM particles (leptons and Higgs bosons). Note that, since we aim to analyze a non-thermal leptogenesis scenario where the reheat temperature 
$T_{RH}<<M_i$, the washout of the asymmetries due to the inverse decay can safely be ignored as the thermal bath does not have sufficient energy to 
reproduce RHNs. Finally, in 
Eq.~(\ref{be}e) we provide the evolution of the number density of the DM fermions ($n_{\psi_{\rm DM}}$). The second term in the RHS of this equation is responsible for the production of DM fermions from the decay of the inflaton, and the third term describes the reduction of the number density of the DM fermions due to their annihilation.
%%%%%%%%%%%%%%%%%%%%%%%%%%%%%%%%%%%%%%%%%%%%
\subsection{Transformation of variables and initial conditions}
In solving the Boltzmann equations, it is useful to use quantities in which the expansion of the Universe is scaled out. Here, we use the following 
transformation of the variables
\beqs
\bea
E_{\phi}&=& \rho_{\phi}a^3,\\
E_{N_i}&=& \rho_{N_i}a^3,\\
N_{L_i}&=& n_{L_i}a^3,\\
R &=&  \rho_{\rm R}a^4,\\
X&=& n_{\psi_{\rm DM}}a^3.
\eea
\label{TV}
\eeqs 
%%%
 We next define the dimensionless quantity $y$ in terms of the scale factor ($a$) and its initial value ($a_I$):
\bea
y\equiv\frac{a}{a_I}.
\eea
The Hubble parameter of the Universe can be written as
\bea
H &=& \sqrt{\frac{1}{3m_{\rm Pl}^2}\bigg(\rho_{\phi}+\sum_{i=1,2,3}\rho_{N_i}+\rho_{\rm R}\bigg)}\nonumber \\
&=& \sqrt{\frac{1}{3m_{\rm Pl}^2a_I^4y^4}\bigg(E_{\phi}a_Iy+\sum_{i=1,2,3}E_{N_i}a_Iy+R\bigg)}.
\label{Hubble}
\eea
%Further, we define a quantity $z$ as
%\bea
%z&=&\frac{M_1}{T}=M_1 a_I y\bigg[\frac{\pi^2}{30g_{*}R}\bigg]^{1/4}
%\eea
%where $T$ can be calculated using the relation, $\rho_R=\frac{\pi^2}{30g_*}T^4$.
Using the rescaled variables, one can rewrite the set of Boltzmann equations in Eq.~(\ref{be}) as
\beqs
\bea
E_{\phi}^{\prime}&=&-\frac{\G_{\phi}}{H}\frac{E_{\phi}}{y},\\
E_{N_i}^{\prime}&=&\frac{1}{Hy}(\G_{\phi\to N_iN_i}E_{\phi}-\G_{N_i}E_{N_i}),\\
N_{L_i}^{\prime}&=&\frac{1}{Hy}\epsilon_i\G_{N_i}N_i,\\
R^{\prime}&=&\frac{1}{H}\sum_{i=1,2,3}\G_{N_i}E_{N_i} a_I,\\
X^{\prime}&=&\frac{1}{Hy}\G_{\phi\to \psi_{\rm DM}\psi_{\rm DM}}\frac{E_{\phi}}{m_{\phi}},
\eea
\label{VT_BE}
\eeqs
where the prime denotes derivation with respect to $y$.

At very early cosmic times, the energy density of the inflaton dominates the Universe with the initial number or energy 
density of the rest of the contents of the Universe being zero. The initial value of the $\phi$ energy density can be 
expressed in terms of the initial expansion rate $H_I$ as $\rho_{\phi_I}=3m_{\rm Pl}^2 H_I^2$. Hence, we set the initial 
values of the variables appearing in Eq.~(\ref{VT_BE}) as
\beq
E_{\phi_I}=3m_{\rm Pl}^2H_I^2a_I,
\label{EI}
\eeq
and
\beq
E_{N_i}=0,~~R_I=0,~~N_{L_i}=0,~~X=0.
\label{IC}
\eeq

Note that since we are interested in exploring non-thermal leptogenesis together with non-thermal production of fermionic 
DM, we must keep in mind that the heavy RHNs ($N_i$) and the DM candidate $\psi_{\rm DM}$ are never in thermal 
equilibrium with the thermal bath ($\rho_{N_i}^{\rm eq}=0$ and $n_{\psi_{\rm DM}}^{\rm eq}=0$). As we are interested in 
quite heavy ($\sim f_a$) DM fermions, we may safely ignore the terms corresponding to DM annihilation in Eq.~(\ref{be}e) 
and also in the production of radiation in Eq.~(\ref{be}c) 
since the DM annihilation cross-sections will be highly suppressed. Such suppressed cross-sections guarantee 
that the pair annihilation of DM remains out of equilibrium $i.e.~ n_{\psi_\text{DM}}\bra\sigma v\ket\lesssim H$ (as was 
also shown and discussed in Ref.~\cite{Lazarides:2020frf}) at any temperature smaller than the reheat temperature and 
hence the DM abundance remains constant. Finally, the direct decay of the inflaton to 
radiation is very small and hence can be neglected in Eq.~(\ref{be}c).
%%%%%%%%%%%%%%%%%%%%%%%%%%%%%%%%%%%%%%%%%%%%%%%%%%%%%%%%%%%%

%%%%%%%%%%%%%%%%%%%%%%%%%%%%%%%%%%%%%%%%%%%%%%%%%%%%%%%%%%%%%

One can express the final primordial lepton asymmetry yields $Y_{L_i}$ in terms of $N_{L_i}$ as
\beq
Y_{L_i}=\frac{n_{L_i}}{s}=\bigg(\frac{45}{2\pi^2g_{*s}}\bigg)\bigg(\frac{\pi^2 g_{*}}{30}\bigg)^{3/4}N_{L_i}R^{-3/4},
\eeq
where $s=(2\pi^2/45)g_{*s}T^3$ represents the entropy density with $T=(30/g_*\pi^2)^{1/4}\rho_R^{1/4}$ being the cosmic 
temperature. Here, $g_*$ ($g_{*s}$) is the effective number of massless degrees of freedom for the energy (entropy) density. 
Finally, the baryon asymmetry yield can be expressed as
\bea
Y_B = -\frac{28}{79}\sum_{i=1,2,3} Y_{L_i}.
\label{bar_asym}
\eea

\noindent Similarly, the fermionic DM relic abundance is
\bea
\Omega_{\psi_{\rm DM}}h^2&=&2\times 2.75\times 10^8\bigg(\frac{m_{\rm DM}}{\rm GeV}\bigg) Y_{\psi_{\rm DM}}\nonumber\\
&=&2.75\times 10^8\bigg(\frac{45}{2\pi^2g_{*s}}\bigg)\bigg(\frac{\pi^2 g_{*}}{30}\bigg)^{3/4}\bigg(\frac{m_{\rm DM}}
{\rm GeV}\bigg)XR^{-3/4},
\eea
where $Y_{\psi_{\rm DM}}=n_{\psi_{\rm DM}}/s$ is the fermionic DM yield. 

\section{Numerical Analysis}
\label{sec:numeric} 

Before delving into the detailed solution of Boltzmann equations, we would like to briefly discuss the possibility of axion as 
a DM candidate in the present model. Clearly, the axion being a integral part of this model can also contribute to the DM relic density, but its 
contribution depends on the choice of the PQ breaking scale $f_a$. The relic axion abundance is expressed as~\cite{Visinelli:2009zm}
\bea
\Omega_ah^2=\Omega_a^{mis}h^2+\Omega_a^{str}h^2\simeq 2.41 \bigg(\frac{f_a}{10^{12}~\text{GeV}}\bigg)^{7/6},
\label{relic_axion_total}
\eea
where $\Omega_a^{mis}h^2$ and $\Omega_a^{str}h^2$ represent the contributions coming from the misalignment 
mechanism~\cite{Preskill:1982cy,Abbott:1982af,Stecker:1982ws,Visinelli:2009zm} and the decay of axionic 
strings~\cite{Hagmann:2000ja,Visinelli:2009zm}, respectively. The relic axion abundance produced by the misalignment mechanism 
is~\cite{Visinelli:2009zm}
\bea
\Omega_ah^2\simeq 0.236 \bigg(\frac{f_a}{10^{12}~\text{GeV}}\bigg)^{7/6}\bra\theta^2f(\theta)\ket,
\eea
where $\theta$ denotes the misalignment angle that lies in the interval $[-\pi,\pi]$~\cite{Dimopoulos:2003ii}. The function 
$f(\theta)$ incorporates the anharmonicity of the axion potential and the mean value $\bra\theta^2f(\theta)\ket$ evaluated in 
the interval $[-\pi,\pi]$ turns out to be around 8.77~\cite{Visinelli:2009zm}. 
On the other hand, the contribution coming from the decay of axionic strings is given by~\cite{Visinelli:2009zm}
\bea
\Omega_a^{str}h^2\simeq 0.34 \bigg(\frac{f_a}{10^{12}~\text{GeV}}\bigg)^{7/6}.
\label{relic_axion_exp_str}
\eea
\begin{figure}[htb!]
	\centering
	\includegraphics[width=0.7\linewidth]{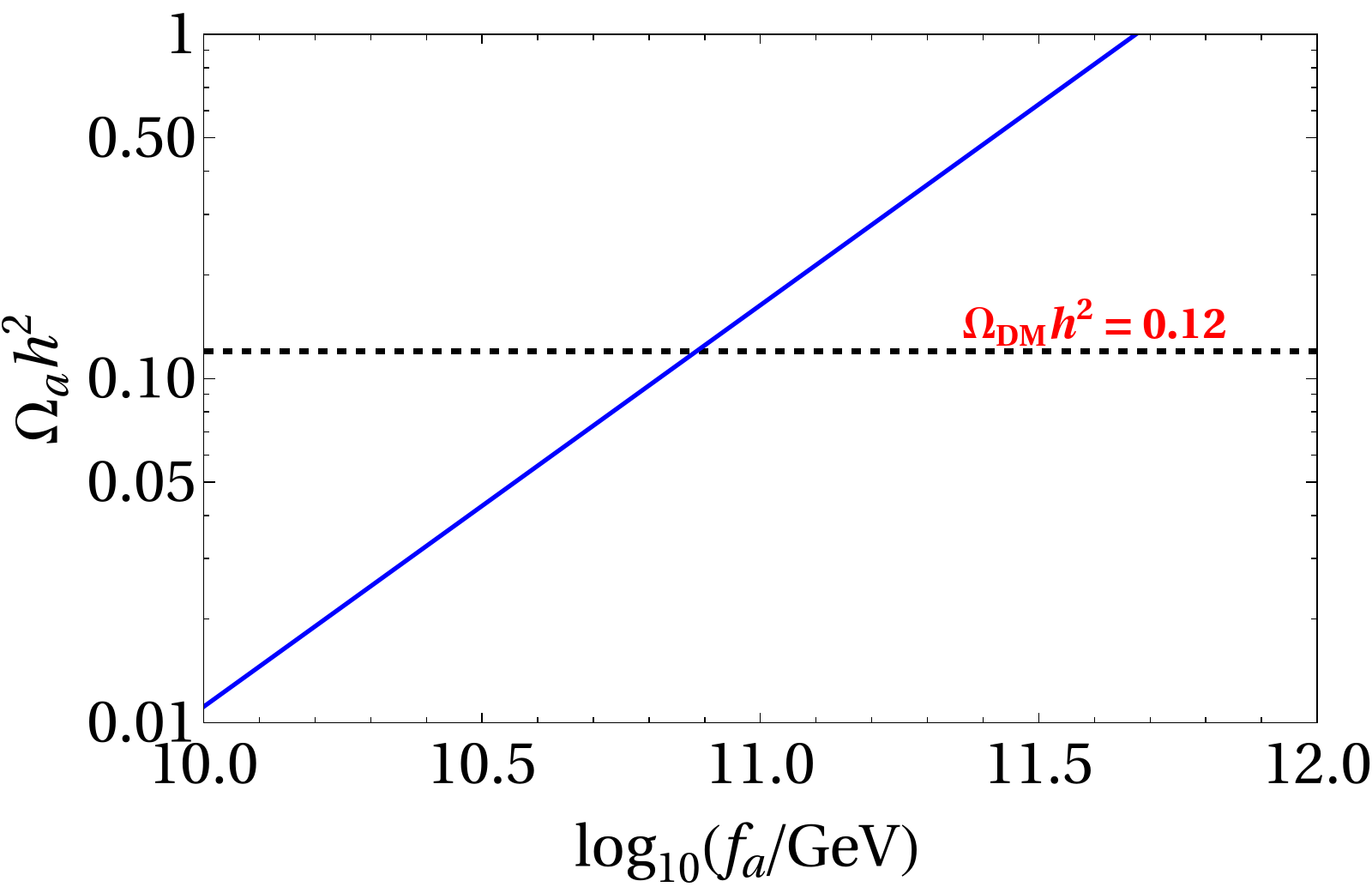}
	\caption{Variation of axion relic abundance with the axion decay constant $f_a$. The black dashed line corresponds to 
      $\Omega_{\text{DM}}h^2=0.12$.}
	\label{relic_axion}
\end{figure} 

\noindent In Fig.~\ref{relic_axion}, we show the variation 
of the axion relic abundance $\Omega_ah^2$ with the axion decay constant $f_a$. The black dashed line corresponds to the Planck 
limit~\cite{Planck:2018vyg} on the total relic DM abundance $\Omega_{\text{DM}}h^2$. Note that with $f_a\simeq8\times 10^{10}$ 
GeV, the axion alone can 
contribute $100\%$ of the relic abundance of DM. The constraint on the relic DM abundance also gives an upper bound on $f_a$, 
suggesting that with $f_a>8\times 10^{10}$ GeV the axion can over-close the energy density of the Universe. The total DM relic 
abundance that should satisfy the Planck limit~\cite{Planck:2018vyg} must include the contribution from the lightest neutral 
component of $\psi_{10}$: 
 \bea
 \Omega_{\rm DM}h^2=\Omega_{a}h^2+\Omega_{\psi_{\rm DM}}h^2.
 \label{relic_t}
 \eea 

Next, we discuss the solution of the Boltzmann equations. To solve them, we need to have information about 
the masses and couplings of the various species involved. To this end, we first demonstrate how we obtain the masses 
of the different RHNs and their corresponding Yukawa couplings. The Lagrangian for the Yukawa interactions of the SM fermions
and RHNs is given by
\begin{align}\label{eq:GUT_Yukawas}
y_{10} \psi_{16}^{\rm T} \mathcal{C} \phi_{10} \psi_{16} + y_{126} \psi_{16}^{\rm T} \mathcal{C} \phi_{126}^\dagger 
\psi_{16} +  {\rm H.c.}  ,
\end{align}
where the transpose $T$ and the charge conjugation operator $\mathcal{C}: {\psi^C}_{L,R} \equiv \mathcal{C}\bar{\psi}_{R,L}^{\rm T}$ 
act in the Dirac space of each left ($L$) or right handed ($R$) fermionic field, and $y_{10}$, $y_{126}$ are $3\times 3$ 
matrices in the family space. The 
VEVs of the four $SU(2)_L$ doublets from the $(2,2,1)$ component of $\phi_{10}$ and 
the $(2,2,15)$ component of $\phi_{126}^\dagger$ are $v_{10}^u$, $v_{10}^d$, $v_{126}^u$, and $v_{126}^d$ with the superscripts $u$ and 
$d$ referring to the up-type and down-type components. The SM Yukawa couplings for the up-and down-type quarks, neutrinos, and charged 
leptons are \cite{Lazarides:1980nt,Babu:1992ia}
\begin{align}\label{eq:SM_Yukawas}
y_u &= \frac{1}{v_{\rm SM}}\left( v_{10}^u y_{10} + v_{126}^u y_{126}\right) , \nonumber \\
y_d &= \frac{1}{v_{\rm SM}}\left( v_{10}^d y_{10} + v_{126}^d y_{126}\right) , \nonumber \\
y_\nu &= \frac{1}{v_{\rm SM}}\left( v_{10}^u y_{10} - 3v_{126}^u y_{126}\right) , \nonumber \\
y_l &= \frac{1}{v_{\rm SM}}\left( v_{10}^d y_{10} - 3v_{126}^d y_{126}\right) ,
\end{align}
where $v_{\rm SM}$ is the SM breaking VEV.
The Majorana mass matrix of the RHNs after the $(1,3,10)\in \overline{126}$ acquires its VEV ($v_R$) at the scale $M_{II}$ is expressed as
\begin{align} \label{eq:M_R}
M_R = y_{126} v_R .
\end{align}
We adopt the parametrization $h\equiv (v_{10}^d/v_{\rm SM}) y_{10}$, $f\equiv (v_{126}^d/v_{\rm SM})y_{126}$, $r \equiv (v_{10}^u/v_{10}^d)$, 
$s\equiv (1/r)(v_{126}^u/v_{126}^d)$, $r_R \equiv v_R (v_{\rm SM}/v_{126}^d)$,
%, and $r_L\equiv (v_{\rm SM}/v_{126}^d)$ 
and re-express the Yukawa couplings as
\begin{align}\label{eq:SM_Yukawas_param}
y_u &= r(h+sf) , \nonumber \\
y_d &= h+f , \nonumber \\
y_\nu &= r(h-3sf) , \nonumber \\
y_l &= h-3f. 
\end{align}
The RHN mass matrix takes the form
\begin{align}
M_R = r_R f .
\end{align}

To estimate the baryon asymmetry of the Universe we need the RHN masses and the Dirac Yukawa couplings ($y_\nu$) of the neutrinos in the 
diagonal basis of the RHN mass matrix ($M_R$) and the charged lepton Yukawa coupling matrix ($y_l$). The fit of the renormalizable 
Yukawa couplings to fermion masses and mixings at the electroweak scale has been extensively studied in several papers 
\cite{Bajc:2005zf,Joshipura:2011nn,Altarelli:2013aqa,Dueck:2013gca,Meloni:2014rga,Meloni:2016rnt,Babu:2016bmy,Ohlsson:2018qpt,
Boucenna:2018wjc,Ohlsson:2019sja, Mummidi:2021anm}. We use the best fit Yukawa couplings 
with the scalars $10_H\oplus \overline{126}_H$ in the $SO(10)\times U(1)_{\rm PQ}$ GUT model from two recent references 
\cite{Ohlsson:2019sja, Mummidi:2021anm}, and we run the fitted 
couplings from the GUT scale ($\sim 10^{16}$ GeV) to the intermediate scale ($\sim 10^{12}$ GeV) using the SM RGEs given in 
Ref.~\cite{Ohlsson:2019sja}. At the scale $10^{12}$ GeV, we diagonalize the RHN mass matrix and compute the Dirac 
Yukawa coupling matrix $y_\nu$ in the same basis where the RHN mass matrix $M_R$ and the charged lepton Yukawa coupling matrix 
$y_l$ are diagonal:
\begin{align}\label{eq:ynu_rel}
M_R^{\rm D} = U_R M_R U_R^{\rm T}, \ y_{l}^{\rm D} = U_l y_l {U_{l}^c}^\dagger\Rightarrow y_\nu \to U_l y_\nu U_R^* .
\end{align}
%%%%%%%%%%%%%%%%%%%%%%%%%%%%%%%%%%%%%%%%%%%%%%%%%%%%%%%%%%%%%%%%%%%%
 
 \subsection {Examples}
\underline{\bf Example I}: The fitted parameters in Ref.~\cite{Ohlsson:2019sja} are
\begin{align}
h&=\begin{pmatrix}
 0.0001 & 0 & 0 \\
 0 & 0.0556 & 0 \\
 0 & 0 & 6.5110 \\
\end{pmatrix}\times 10^{-3} ,
\nonumber \\
f&=\begin{pmatrix}
 0.00556-0.00318 i & -0.01130-0.01208 i & -0.03546-0.14594 i \\
 -0.01130-0.01208 i & -0.16392+0.03471 i & -0.25650+0.25582 i \\
 -0.03546-0.14594 i & -0.25650+0.25582 i & -0.91962-0.50277 i 
\end{pmatrix}\times 10^{-3} , \nonumber \\
r&=-65.9350{\rm ,} \ s = 0.391447-9.585\times 10^{-17} i {\rm ,} \ r_R = 2.04454\times 10^{15} \ {\rm GeV.}
\end{align}
%At the scale $10^{12}$ GeV, we have diagonalized the RHN mass matrix and have computed the Dirac Yukawa coupling matrix $y_\nu$ 
%in the same basis where the RHN mass matrix $M_R$ and the lepton Yukawa coupling matrix $y_l$ are diagonal.
The physical masses of the RHNs are obtained as
\begin{align}
M_R^D=\left\{1.87\times 10^{10},4.46\times 10^{11},2.36\times 10^{12}\right\} \ \mathrm{GeV},
\end{align}
and the Dirac Yukawa structure of the light neutrinos in the basis described in Eq.~(\ref{eq:ynu_rel}) is given by
\begin{align}
U_l y_\nu U_R^* = \begin{pmatrix}
 -0.00123+0.00077 i & 0.00261+0.00204 i & -0.00258+0.01029 i \\
 -0.00229-0.00012 i & -0.00629+0.01499 i & -0.00719+0.03334 i \\
 -0.05876+0.03800 i & 0.08977+0.07709 i & -0.15355+0.45175 i 
\end{pmatrix}  .
\end{align}

Next, we plug these values of the RHN masses and the Yukawa couplings into the Boltzmann equations  (Eq.~(\ref{VT_BE})) and we also 
fix $m_{\phi}=1.7\times 10^{13} ~\text{GeV}$ corresponding to the parameter values for successful inflation in 
Section~\ref{sec:inflation} \cite{Lazarides:2020frf} and $m_{126} = 7.5\times 10^{12}$ GeV. These values of $m_{\phi}$ and 
$m_{126}$ together with the given RHN masses are chosen to obtain the baryon asymmetry's desired value. 
 Form  Fig.~\ref{relic_axion} one sees that depending on the choice of the axion decay constant $f_a$, the axion can fully account 
for or partially contribute to the DM relic abundance. With this in mind, we set $f_a\simeq m_{45}=5\times10^{10}$ GeV, such that the 
axion contributes $\sim 60\%$ ($\Omega_{a}h^2 = 0.073$) of the DM relic abundance, and the remaining $40\%$ comes from the fermionic 
component. 
%%%%%%%%%%%%%%%%%%%%%%%%%%%%%%%%%%%%%%%%%%%%%%%%%%%%%%%%%%
  \begin{figure}[htbp!]
 	\centering
 	\includegraphics[width=0.7\linewidth]{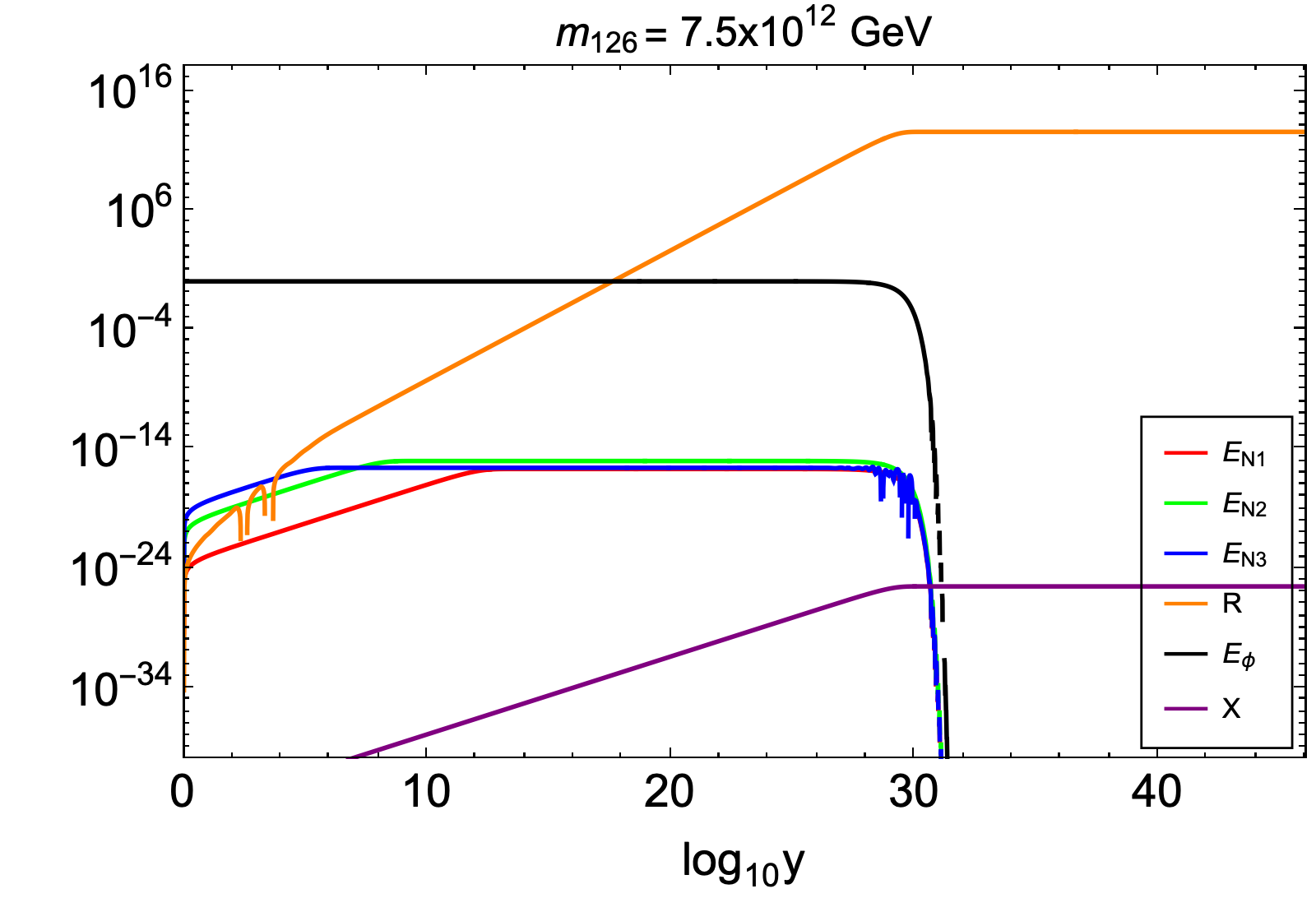}
 	\caption{Evolution of the rescaled energy and number densities of the various components in the Universe (normalized to $E_{\phi_I}$ ) 
versus $\log_{10}y$ for fixed values of $m_{\phi}=1.7\times10^{13}~\text{GeV}$ and $m_{126} = 7.5\times 10^{12}$ GeV.  Here we have set 
 $m_{\rm DM}=7\times 10^{9}$ GeV and $m_{45}=5\times 10^{10}$ GeV such that $\Omega_{\psi_{\rm DM}}h^2=0.046$.}
 	\label{yield_inf_ex1}
 \end{figure} 
%%%%%%%%%%%%%%%%%%%%%%%%%%%%%%%%%%%%%%%%%%%%%%%%%%%%%%%%%%%%%% 

%%%%%%%%%%%%%%%%%%%%%%%%%%%%%%%%%%%%%%%%%%%%%%%
\begin{figure}[h!]
	\centering
	\subfloat[]{\includegraphics[width=0.48\linewidth]{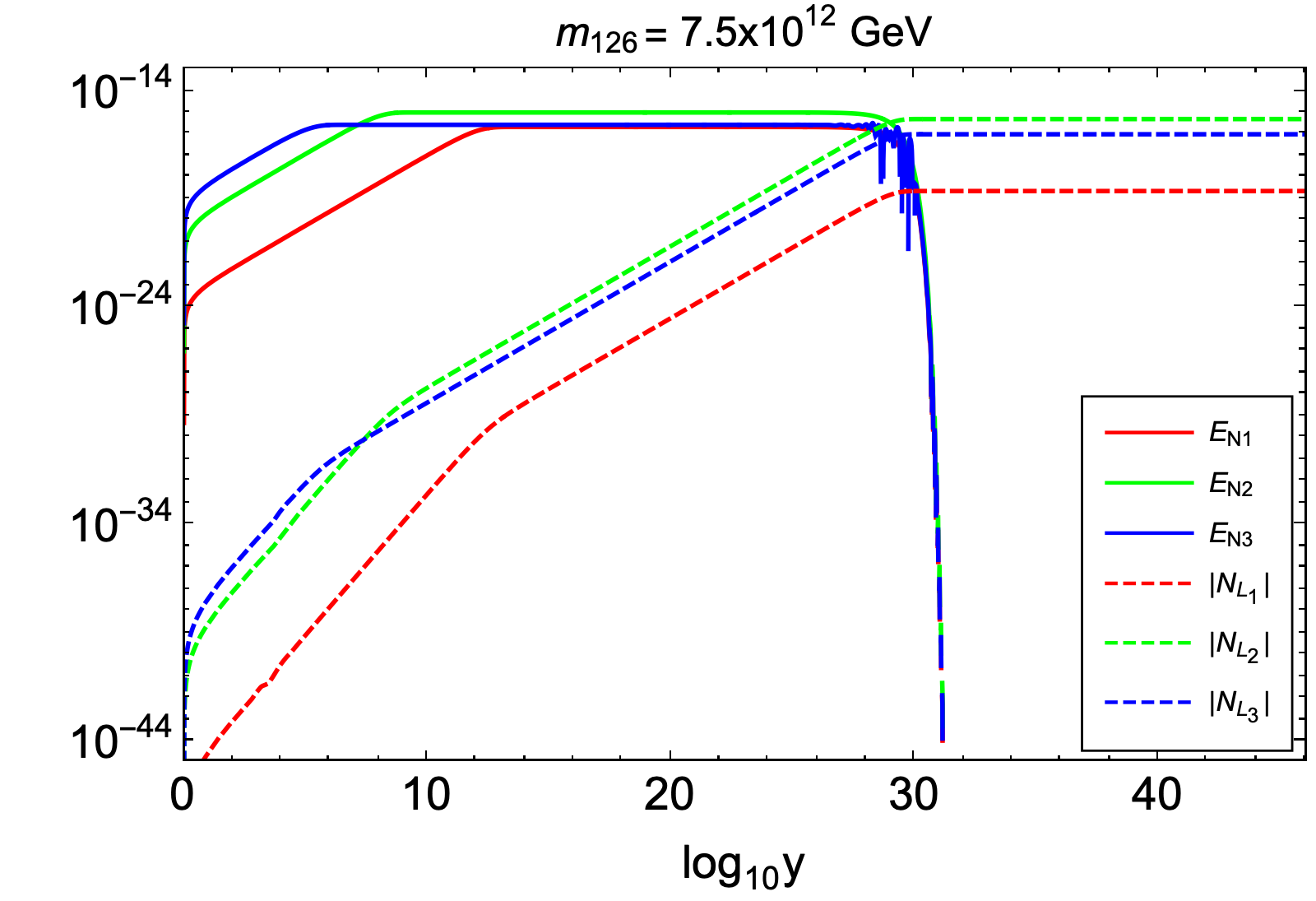}} \hspace{4mm}
	\subfloat[]{\includegraphics[width=0.48\linewidth]{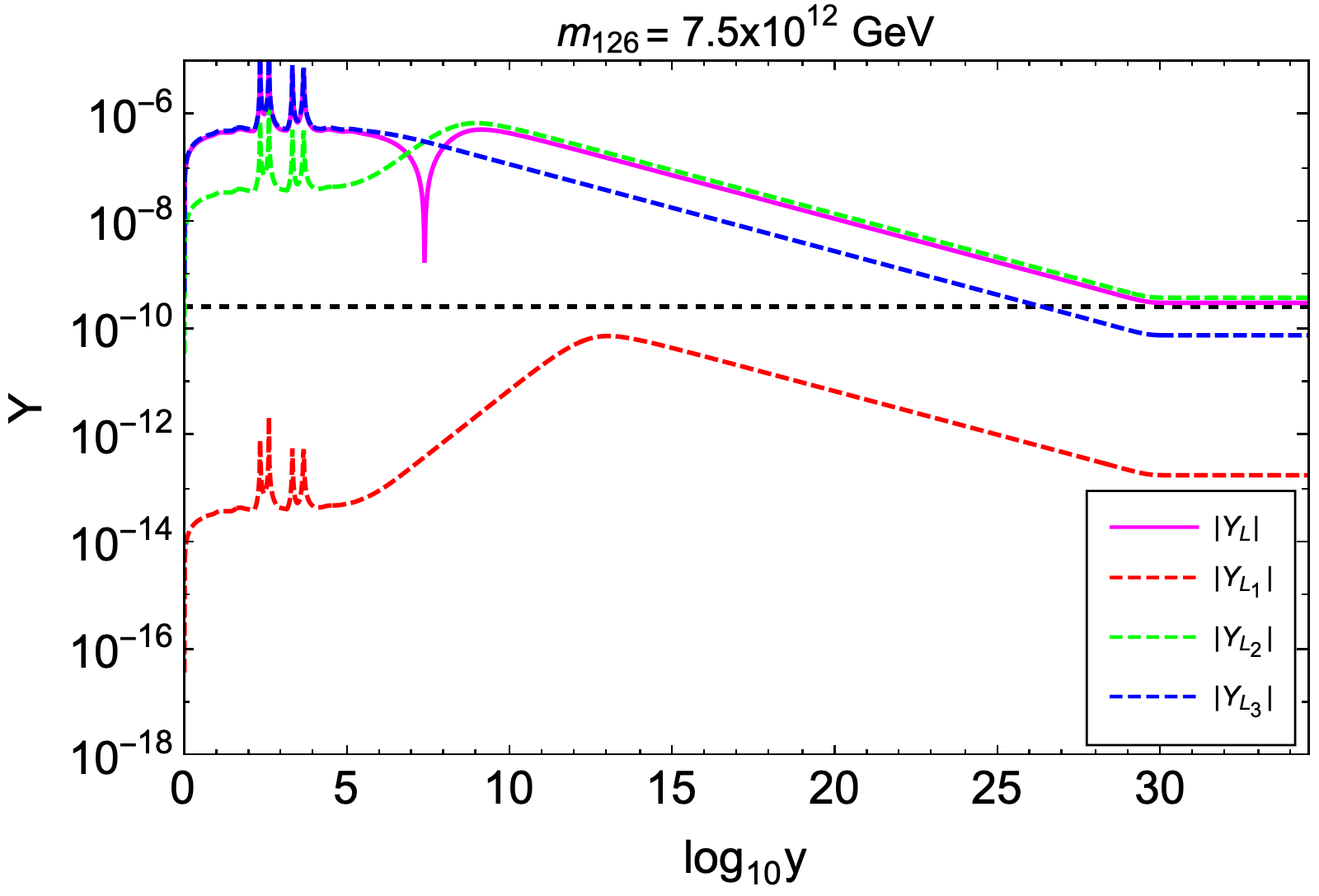}}\\
	\subfloat[]{\includegraphics[width=0.48\linewidth]{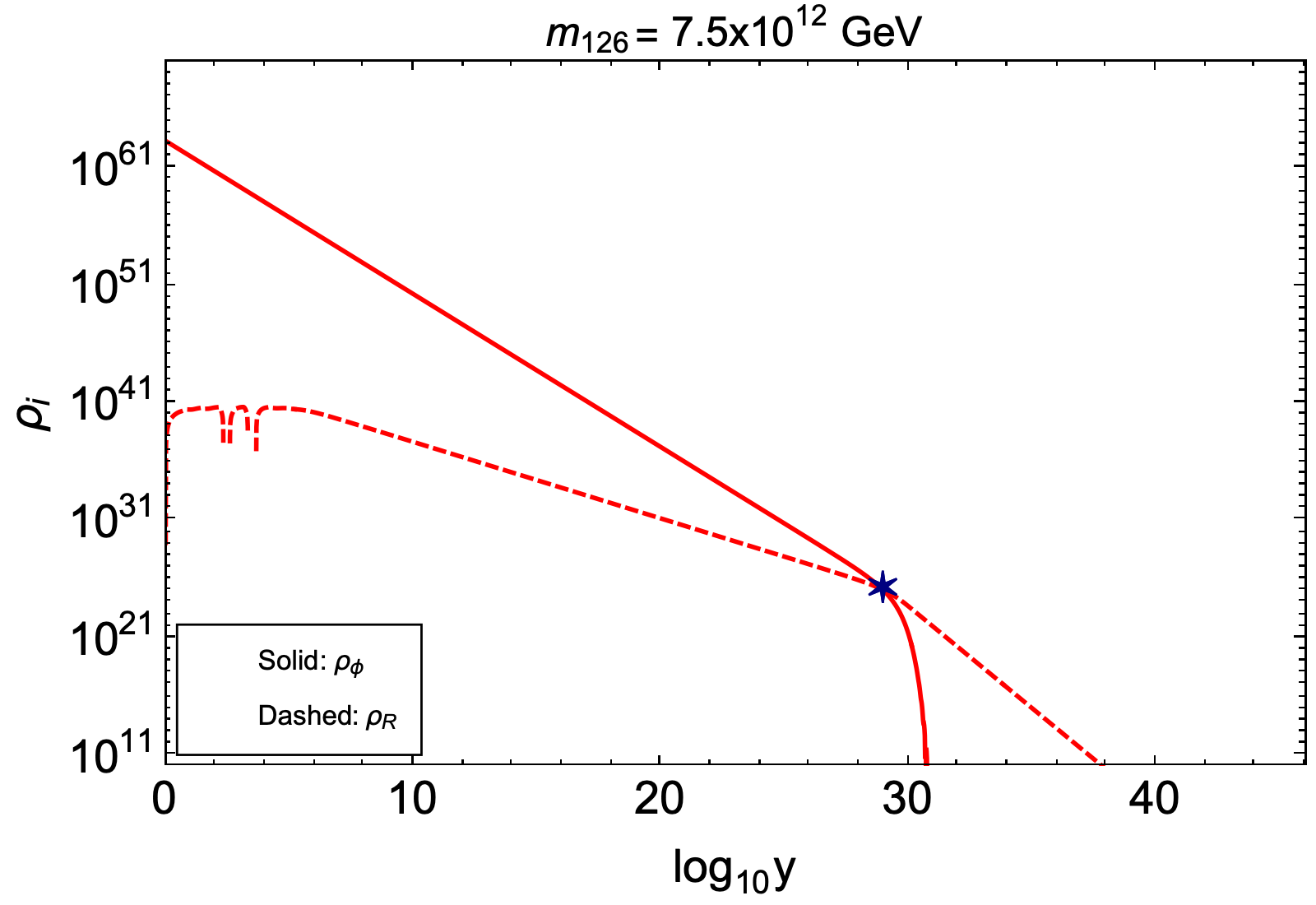}} \hspace{4mm}
	\subfloat[]{\includegraphics[width=0.48\linewidth]{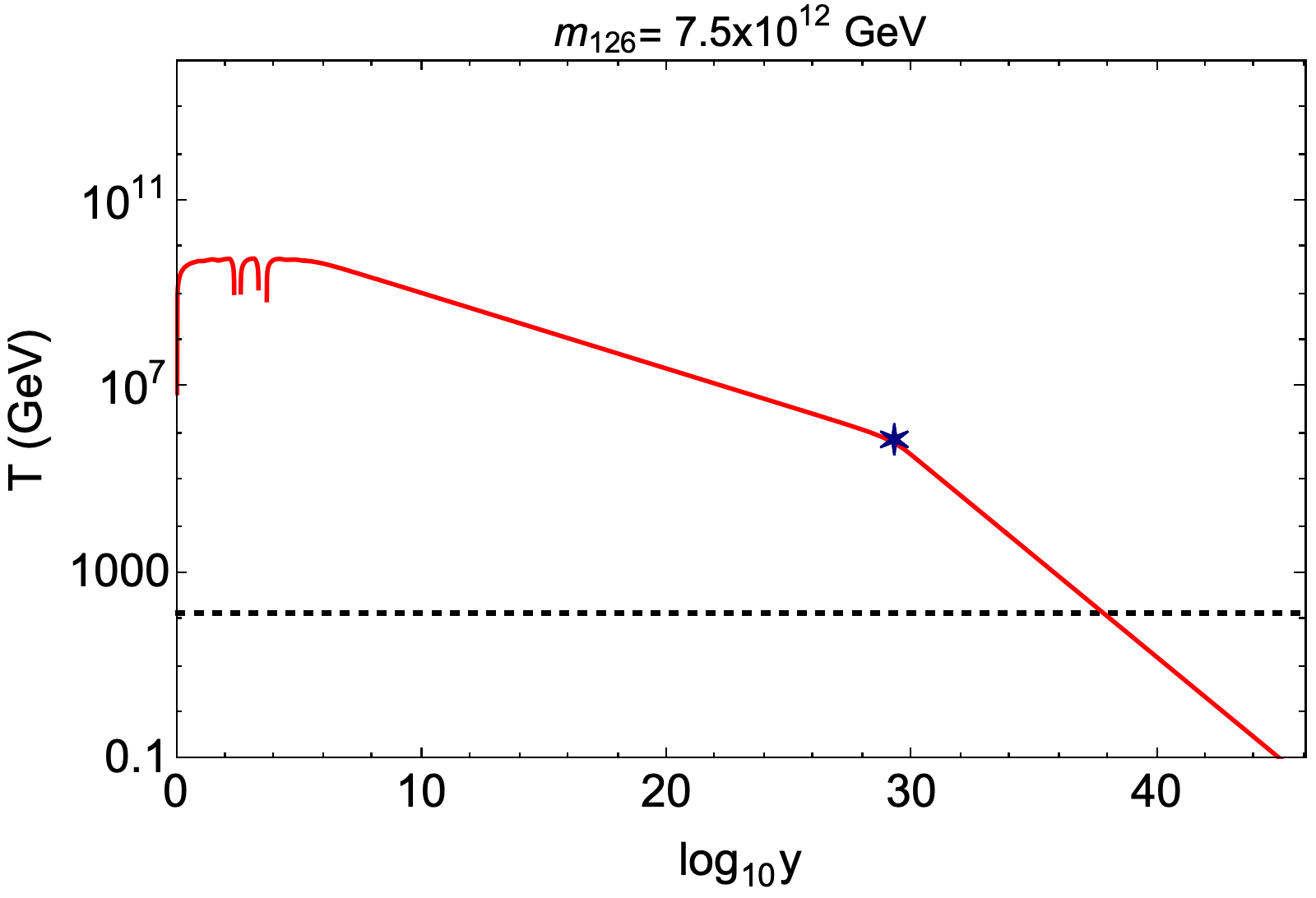}}
	\caption{Top left panel: evolution of the rescaled energy densities of all three RHNs together with the corresponding lepton asymmetries 
(normalized to $E_{\phi_I}$ ) as functions of $\log_{10}y$. Top right plot: evolution of the individual lepton asymmetry yields together 
with the total asymmetry. The horizontal dashed line corresponds to the total primordial lepton asymmetry ($Y_L\simeq 2.5\times 10^{-10}$) that 
gives the observed baryon asymmetry of the Universe ($Y_B\simeq 8.7\times10^{-11}$). Bottom left panel: evolution of $\rho_\phi $ 
and $\rho_R$ with $\log_{10}y$. The $\star$ mark shows the cross-over of $\rho_\phi $ and $\rho_R$. Bottom right panel: evolution of the 
bath's temperature with $\log_{10}y$. Here the $\star$ mark corresponds to $T_{RH}$ and the horizontal dashed line shows the sphaleron 
free-out temperature $T_{sph}\simeq 130$ GeV. All these plots are made for a fixed value of $m_{\phi}=1.7\times10^{13}~\text{GeV}$ and  
$m_{126} = 7.5\times 10^{12}$ GeV.}
	\label{asymmetry_ex1}
\end{figure}
%%%%%%%%%%%%%%%%%%%%%%%%%%%%%%%%%%%%%%%%%%%%%%%%%%%%%%%%%%%%%%%
 
Solving the Boltzmann equations we find the evolution of the different species, which are crucial for 
baryon asymmetry and the DM abundance in the Universe, as functions of the dimensionless quantity $y=a/a_I$ where we have set 
$a_I = 1$~\cite{Hahn-Woernle:2008tsk}. The results are depicted in Fig.~\ref{yield_inf_ex1}. All these rescaled energy and number densities are normalized by the 
initial energy density of the inflaton $E_{\phi_I}$. Here, the black line corresponds to the inflaton, and the red, green, and 
blue lines to the RHNs $N_1$, $N_2$, and $N_3$ respectively. The orange and purple lines respectively correspond to the radiation and the 
fermionic DM. One notices that the rescaled inflaton energy density initially remains constant but, at $\log_{10}y\simeq30$, completely 
decays to RHNs and fermionic DM. As expected, the fermionic DM rescaled number density slowly 
increases as a result of its production from the inflaton decay until the point when the decay of the inflaton is completed. Thereafter, 
it remains constant giving rise to the final fermionic DM abundance that corresponds to almost $40\%$ of the DM relic abundance of the 
Universe for $m_{\rm DM} = 7\times 10^{9}$ GeV. The RHN rescaled energy densities also gradually increase due to their production by the 
inflaton decay and are subsequently stabilized when their production rate from the inflaton decay becomes comparable 
to their decay rate to SM particles. Being the heaviest among all RHNs, $N_3$ has the largest decay width ($\Gamma_{N_3}\propto M_3$). 
Consequently, the rescaled energy density of $N_3$ is stabilized much earlier than that of the other two RHNs. It is interesting to point out that 
the decay of the RHNs is almost completed at the same time as that of the inflaton. This is because the RHNs decay instantaneously 
after their production as $\Gamma_{N_i}>>\Gamma_\phi$ in all cases. As a result of the RHN decay, a rise in the rescaled radiation energy density  
is also observed. The production of radiation stops once all the RHNs have completely decayed.
 
  In the top left panel of Fig.~\ref{asymmetry_ex1}, we show the evolution of the rescaled energy densities of the RHNs and the lepton 
asymmetries (scaled by $E_{\phi_I}$) generated by their CP-violating decays. The behavior of these asymmetries is similar to that of radiation in 
Fig.~\ref{yield_inf_ex1}. In the top right panel of Fig.~\ref{asymmetry_ex1}, we plot the absolute values of the lepton asymmetry 
yields ($Y_{L_i}=n_{L_i}/s$) generated by the decay of the RHNs together with the absolute value of the total lepton asymmetry yield 
(shown in magenta) which leads, for large values of $y$, to the primordial lepton asymmetry (shown by the black dashed line) required 
to produce the observed baryon asymmetry of the Universe. The dip observed in the total lepton asymmetry $Y_L$ is a result of 
cancellation due to the different signs of the individual lepton asymmetries generated in the decay of the RHNs. The decrease of these 
asymmetries after some point is caused by the entropy injection into the bath. This happens when the decay rates of the RHNs become 
comparable to their production rates and as a result, the radiation starts to build at a faster rate as can also be seen from 
Fig.~\ref{asymmetry_ex1}. Finally, the asymmetries stabilize once the decay of the RHNs is complete. 
 In the bottom left panel of Fig.~\ref{asymmetry_ex1} we show the evolution of the energy densities of the inflaton $\rho_\phi$ 
(red solid) and the radiation $\rho_R$ (red dashed). This plot shows that the radiation-dominated era starts only when 
$\rho_R=\rho_\phi$ is reached, and we use this condition to determine the Universe's reheat temperature ($T_{RH}$). In the 
bottom right plot of Fig.~\ref{asymmetry_ex1} we show the evolution of the thermal bath's temperature given by
\bea
T_R=\bigg(\frac{30}{g_*\pi^2}\bigg)^{1/4}\rho_R^{1/4},
\label{T_bath}
\eea
where $g_*=106$. For the above choices of parameters, the reheat temperature of the Universe is found to be $T_{RH} = 1.9\times 10^6$ GeV 
(shown by the $\star$). Here, we also show a black dashed horizontal line that corresponds to the sphaleron freeze-out temperature 
$T_{sph}=130$ GeV~\cite{Kuzmin:1985mm,Bento:2003jv,DOnofrio:2014rug}.

\underline{\bf Example II}: The fitted parameters in Ref.~\cite{Mummidi:2021anm} are
\begin{align}
h&=\begin{pmatrix}
 0.00023 & 0 & 0 \\
 0 & -0.04811 & 0 \\
 0 & 0 & -5.79504 \\
\end{pmatrix}\times 10^{-3} ,
\nonumber \\
f&=\begin{pmatrix}
 -0.00088+0.00178 i & 0.00475-0.00889 i & 0.04635+0.06797 i \\
 0.00475-0.00889 i & 0.11279+0.05108 i & -0.12218-0.25921 i \\
 0.04635+0.06797 i & -0.12218-0.25921 i & 0.54683-0.59856 i \\
\end{pmatrix}\times 10^{-3} , \nonumber \\
r&=77.4189{\rm ,} \ s = 0.314-0.0282 i {\rm ,} \ r_R = 9.84\times 10^{14} \ {\rm GeV.}
\end{align}
At the scale $10^{12}$ GeV, we have diagonalized the RHN mass matrix and computed the Dirac 
Yukawa matrix $y_\nu$ in the basis where the RHN mass matrix ($M_R$) and the charged lepton Yukawa coupling 
matrix ($y_l$) are diagonal.
%%%%%%%%%%%%%%%%%%%%%%%%%%%%%%%%%%%%%%%%%%%%%%%
\begin{figure}[h!]
	\centering
	\subfloat[]{\includegraphics[width=0.48\linewidth]{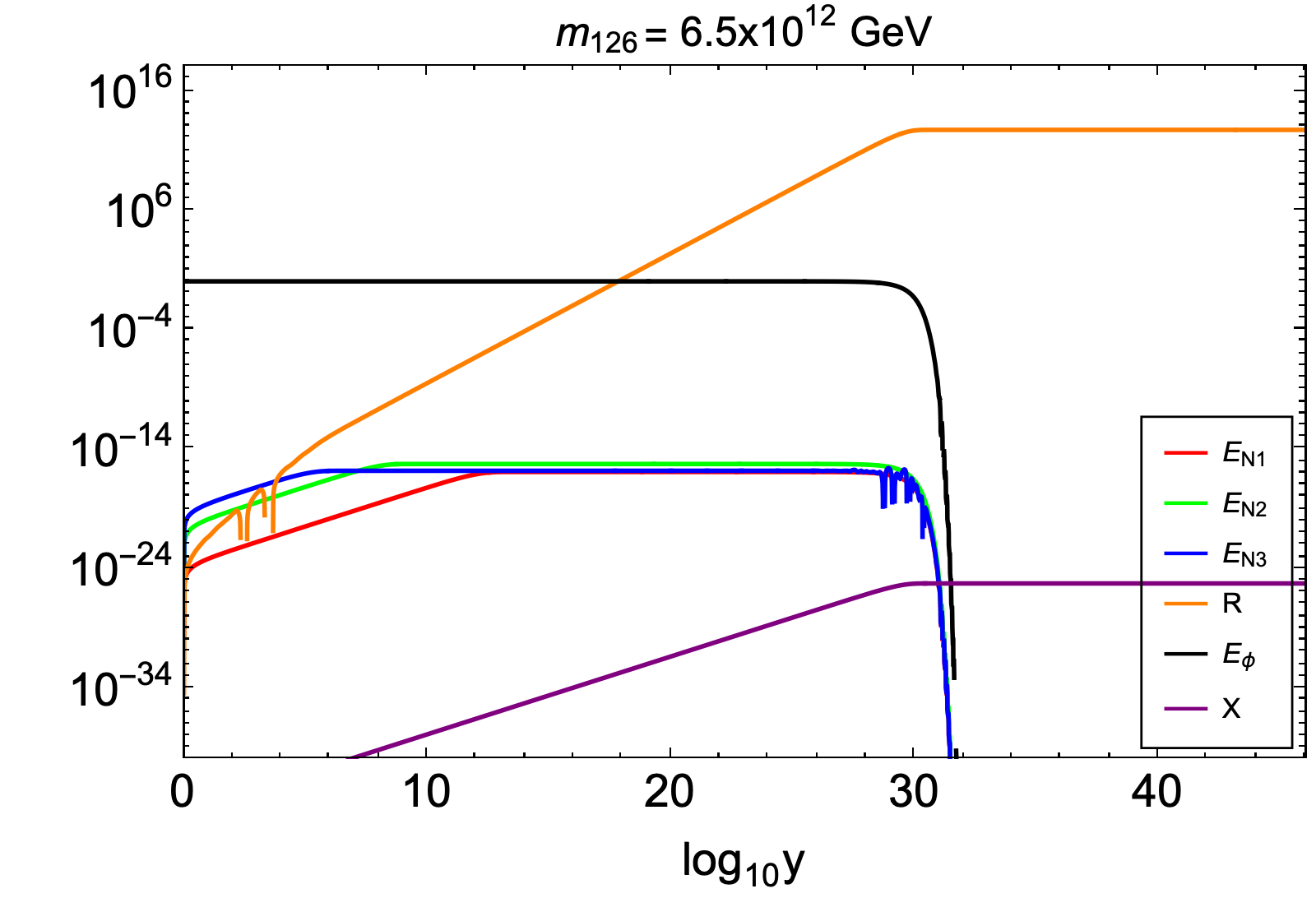}} \hspace{4mm}
	\subfloat[]{\includegraphics[width=0.48\linewidth]{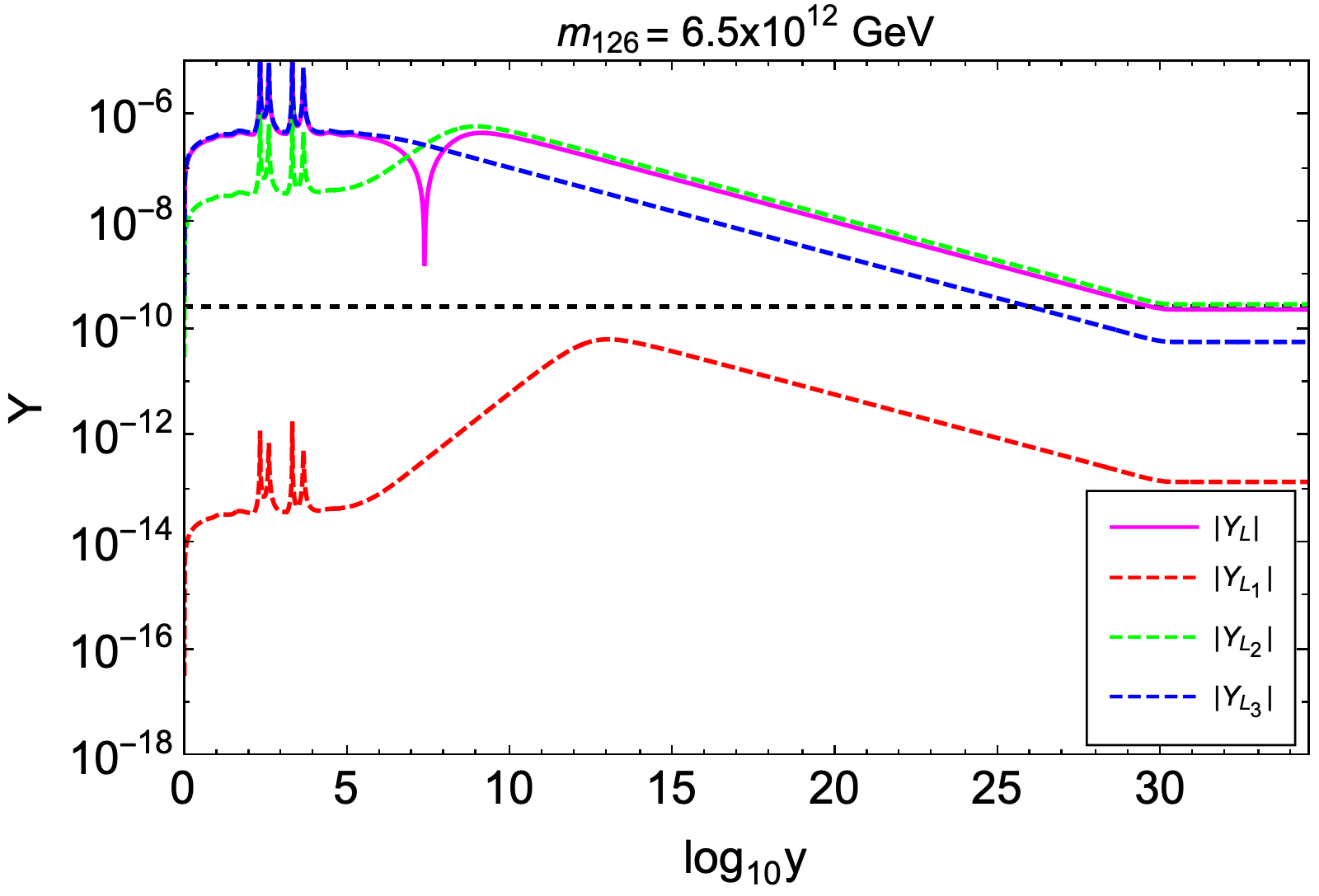}}\\
	\caption{Left panel: evolution of the rescaled energy and number densities of various components (normalized to 
$E_{\phi_I}$ ) as functions of  $\log_{10}y$ for fixed values of  $m_{\phi}=1.7\times10^{13}~\text{GeV}$ and  
$m_{126} = 6.5\times 10^{12}$ GeV. Here we have set $m_{\rm DM}=4.5\times 10^{9}$ GeV and $m_{45}=5\times 10^{10}$ GeV 
such that $\Omega_{\psi_{\rm DM}}h^2=0.041$. Consequently, the fermionic DM contributes $34\%$ of the DM in the Universe. 
Right panel: evolution of the individual lepton asymmetry yields together with the total asymmetry. 
The horizontal dashed line corresponds to the total lepton asymmetry ($Y_L\simeq 2.5\times10^{-10}$) that gives the 
observed baryon asymmetry ($Y_B\simeq 8.7\times10^{-11}$).}
\label{asymmetry_ex2}
\end{figure}
%%%%%%%%%%%%%%%%%%%%%%%%%%%%%%%%%%%%%%%%%%%%%%%%%%%
 
 The physical masses of the RHNs are obtained as
\begin{align}
\left\{4.36\times 10^9,1.97\times 10^{11},8.66\times 10^{11}\right\} \ \mathrm{GeV},
\end{align}
and the Dirac Yukawa structure of the neutrinos in the basis of Eq.~(\ref{eq:ynu_rel}) is given by
\begin{align}
 U_l y_\nu U_R^*= \begin{pmatrix}
 -0.00009-0.00081 i & 0.00016+0.00261 i & -0.00841+0.00328 i \\
 0.00077+0.00137 i & -0.00492+0.00440 i & 0.03049-0.00814 i \\
 -0.00667-0.03632 i & 0.02004+0.11319 i & -0.40445+0.22776 i 
\end{pmatrix}  .
\end{align}
Following the previous analysis in Example I we present the results in Fig.~\ref{asymmetry_ex2}. 

%%%%%%%%%%%%%%%%%%%%%%%%%%%%%%%%%%%%%%%%%%%%%%%%%%%%%%%%%%%%%%%%%%%%%%%%
The behavior of the system in this second example is, for all practical
purposes, identical to its behavior in Example I. The axion decay constant in this case is 
$f_a\simeq 5.5\times 10^{10}$~GeV
and the axion contributes about $66\%$ of the DM in the Universe ($\Omega_{a}h^2=0.079$).

\section{Conclusions}
\label{sec:conclusion}

We reconsidered the non-supersymmetric $SO(10)\times U(1)_{\text{PQ}}$ model of Ref.~\cite{Lazarides:2020frf}, but with two 
intermediate gauge symmetry breaking scales and discussed in detail the gauge coupling unification and proton decay by 
taking into account the effect of threshold corrections and dimension-5 operators. We also performed a detailed analysis of 
the DM phenomenology and matter-antimatter asymmetry of the Universe by solving a set of coupled Boltzmann equations. The 
DM in this model is a mixture of axions and an intermediate scale 
mass fermion which is produced non-thermally via the decay of a gauge singlet scalar field with a Coleman-Weinberg potential that 
plays the role of the inflaton. The inflaton also decays to the three RHNs in the $SO(10)$ matter 16-plets. 
Their subsequent decay into SM particles is responsible for generating a primordial lepton asymmetry which, with the help of 
electroweak sphaleron effects, provides the observed baryon asymmetry in the Universe. The model predicts the existence 
of intermediate scale magnetic monopoles with a possibly measurable flux. It also generates intermediate scale
cosmic strings which produce a stochastic gravitational background that can be detected 
in the present and ongoing gravitational wave experiments. 
%%%%%%%%%%%%%%%%%%%%%%%%%%%%%%%%%%%%%%%%%%%%%%%%%%%%%%%%%%%%%%%%%%%%%%%%
\section*{Acknowledgements}
This work is supported by the Hellenic Foundation for Research and Innovation (H.F.R.I.) under
the “First Call for H.F.R.I. Research Projects to support Faculty Members and Researchers and
the procurement of high-cost research equipment grant” (Project Number: 2251). R.R. also
acknowledges the National Research Foundation of Korea (NRF) grant funded by the Korea
government (NRF-2020R1C1C1012452).
%%%%%%%%%%%%%%%%%%%%%%%%%%%%%%%%%%%%%%%%%%%%%%%%%%%%%%%%%%%%%%%%%%%%%%%%%%%%%%%%%%%%%%%%%%%%
\appendix
\section*{Appendix: Threshold Corrections}\label{appen:thr_cor}
\begin{align} \label{eq:thr_cor_mU}
\lambda_{2L}(M_U) & = 30 \log \frac{ m_S\left({3, 1, 15}\right)}{M_U}+ 20 \log \frac{ m_S\left({2, 2, 10}\right)}{M_U}+ 40 \log \frac{ m_S\left({3, 1, 10}\right)}{M_U}\nonumber \\ & + 12 \log \frac{ m_S\left({2, 2, 6}\right)}{M_U}+ 4 \log \frac{ m_S\left({3, 1, 1}\right)}{M_U},\nonumber \\
 \lambda_{2R}(M_U) & = 30 \log \frac{ m_S\left({1, 3, 15}\right)}{M_U}+ 20 \log \frac{ m_S\left({2, 2, 10}\right)}{M_U}+ 12 \log \frac{ m_S\left({2, 2, 6}\right)}{M_U},\nonumber \\
 \lambda_{4C}(M_U) & = 12 \log \frac{ m_S\left({3, 1, 15}\right)}{M_U}+ 12 \log \frac{ m_S\left({1, 3, 15}\right)}{M_U}+ 24 \log \frac{ m_S\left({2, 2, 10}\right)}{M_U}\nonumber \\ & + 18 \log \frac{ m_S\left({3, 1, 10}\right)}{M_U}+ 8 \log \frac{ m_S\left({2, 2, 6}\right)}{M_U}+ 2 \log \frac{ m_S\left({1, 1, 6}\right)}{M_U}. 
 \end{align}
\begin{align} \label{eq:thr_cor_mI}
\lambda_{2L}(M_{I}) & = 6 \log \frac{ m_S\left({2, 2, \overline{3}, -\frac{4}{3}}\right)}{M_{I}}+ 6 \log \frac{ m_S\left({2, 2, 3, \frac{4}{3}}\right)}{M_{I}}+ 16 \log \frac{ m_S\left({2, 2, 8, 0}\right)}{M_{I}},\nonumber \\
 \lambda_{2R}(M_{I}) & =\log \frac{ m_S\left({1, 1, 8, 0}\right)}{M_{I}}+ 12 \log \frac{ m_S\left({1, 3, \overline{3}, \frac{2}{3}}\right)}{M_{I}}+ 24 \log \frac{ m_S\left({1, 3, \overline{6}, -\frac{2}{3}}\right)}{M_{I}}\nonumber \\ & + 4 \log \frac{ m_S\left({2, 2, \overline{3}, -\frac{4}{3}}\right)}{M_{I}}+ 4 \log \frac{ m_S\left({2, 2, 3, \frac{4}{3}}\right)}{M_{I}}+ 16 \log \frac{ m_S\left({2, 2, 8, 0}\right)}{M_{I}}\nonumber \\ & + 2 \log \frac{ m_S\left({1, 1, 8, 0}\right)}{M_{I}},\nonumber \\
 \lambda_{3C}(M_{I}) & = 3 \log \frac{ m_S\left({1, 1, 8, 0}\right)}{M_{I}}+ 3 \log \frac{ m_S\left({1, 3, \overline{3}, \frac{2}{3}}\right)}{M_{I}}+ 15 \log \frac{ m_S\left({1, 3, \overline{6}, -\frac{2}{3}}\right)}{M_{I}}\nonumber \\ & + 4 \log \frac{ m_S\left({2, 2, \overline{3}, -\frac{4}{3}}\right)}{M_{I}}+ 4 \log \frac{ m_S\left({2, 2, 3, \frac{4}{3}}\right)}{M_{I}}+ 24 \log \frac{ m_S\left({2, 2, 8, 0}\right)}{M_{I}}\nonumber \\ & +\log \frac{ m_S\left({1, 1, \overline{3}, -\frac{4}{3}}\right)}{M_{I}}+\log \frac{ m_S\left({1, 1, 3, \frac{4}{3}}\right)}{M_{I}}+ 6 \log \frac{ m_S\left({1, 1, 8, 0}\right)}{M_{I}},\nonumber \\
 \lambda_{B-L}(M_{I}) & = 8 \log \frac{ m_S\left({1, 3, \overline{3}, \frac{2}{3}}\right)}{M_{I}}+ 16 \log \frac{ m_S\left({1, 3, \overline{6}, -\frac{2}{3}}\right)}{M_{I}}+ \frac{128}{3} \log \frac{ m_S\left({2, 2, \overline{3}, -\frac{4}{3}}\right)}{M_{I}}\nonumber \\ & + \frac{128}{3} \log \frac{ m_S\left({2, 2, 3, \frac{4}{3}}\right)}{M_{I}}+ \frac{32}{3} \log \frac{ m_S\left({1, 1, \overline{3}, -\frac{4}{3}}\right)}{M_{I}}+ \frac{32}{3} \log \frac{ m_S\left({1, 1, 3, \frac{4}{3}}\right)}{M_{I}}. 
 \end{align} 
\begin{align} \label{eq:thr_cor_mII}
\lambda_{2L}(M_{II}) & =\log \frac{ m_S\left({2, \frac{1}{2}, 1}\right)}{M_{II}}+\log \frac{ m_S\left({2, \frac{1}{2}, 1}\right)}{M_{II}}+\log \frac{ m_S\left({2, -\frac{1}{2}, 1}\right)}{M_{II}},\nonumber \\
 \lambda_{Y}(M_{II}) & = 8 \log \frac{ m_S\left({1, 2, 1}\right)}{M_{II}}+ 2 \log \frac{ m_S\left({1, 1, 1}\right)}{M_{II}}+ 2 \log \frac{ m_S\left({1, -1, 1}\right)}{M_{II}}\nonumber \\ &+\log \frac{ m_S\left({2, \frac{1}{2}, 1}\right)}{M_{II}}+\log \frac{ m_S\left({2, \frac{1}{2}, 1}\right)}{M_{II}}+\log \frac{ m_S\left({2, -\frac{1}{2}, 1}\right)}{M_{II}},\nonumber \\
 \lambda_{3C}(M_{II}) & =0. 
 \end{align} 
%%%%%%%%%%%%%%%%%%%%%%%%%%%%%%%%%%%%%%%%%%%%%%%%%%%%%%%%%%%%%%%%%%%%
\bibliographystyle{mystyle}
\bibliography{GUT_TD}
%%%%%%%%%%%%%%%%%%%%%%%%%%

\end{document}